\documentclass[a4paper,fleqn,usenatbib]{mnras}

\usepackage{newtxtext,newtxmath}

\usepackage[T1]{fontenc}
\usepackage{ae,aecompl}
\usepackage{mathrsfs}

\usepackage{graphicx}	
\usepackage{amsmath}	
\usepackage{amssymb}	
\usepackage[flushleft]{threeparttable}	
\usepackage{color}

\title[A Fabry-Perot with adaptive optics ]{First results from SAM-FP: Fabry-Perot observations with ground layer adaptive optics -- The structure and kinematics of the core of 30 Doradus }
\author[C. Mendes de Oliveira et al.]{
C. Mendes de Oliveira,$^{1}$\thanks{E-mail: claudia.oliveira@iag.usp.br}
P. Amram,$^{2}$
Bruno C. Quint,$^{3}$
S. Torres-Flores,$^{4}$  \\
\newauthor{R. Barb\'a$^{4}$ and D. Andrade$^{1}$}
\\
$^{1}$
Instituto de Astronomia, Geof\'isica e Ci\^encias Atmosf\'ericas da U. de S\~{a}o Paulo, Cidade Universit\'aria, 
05508-900, S\~{a}o Paulo, SP, Brazil \\
$^{2}$ Aix Marseille Universit\'e, CNRS, LAM (Laboratoire d'Astrophysique de Marseille), Marseille, France \\
$^{3}$ SOAR Telescope, AURA-O, Colina El Pino S/N, Casila 603, La Serena, Chile \\  
$^{4}$ Departamento de F\'isica y Astronom\'ia, Universidad de La Serena, Av. Cisternas 1200, La Serena, Chile
}

\date{Accepted XXX. Received YYY; in original form ZZZ}

\pubyear{2016}

\begin{document}
\label{firstpage}
\pagerange{\pageref{firstpage}--\pageref{lastpage}}
\maketitle

\begin{abstract}

The aim of this paper is to present the first data set obtained with SAM-FP, a Fabry-Perot instrument mounted inside the SOAR telescope Adaptive-Optics Module. This is the only existing imaging Fabry-Perot interferometer using laser-assisted ground layer adaptive optics. SAM-FP was used to observe the ionised gas, traced by H$\alpha$, in the centre of the 30 Doradus starburst (the Tarantula Nebula) in the LMC, with high spatial ($\sim$0.6
arcsec, or 0.15 pc) and spectral ($ R \simeq 11200 $) resolution. Radial velocity, velocity dispersion and monochromatic maps were derived. The region displays a mix of narrow, $\sigma$ $\sim$ 20 km
s$^{-1}$ profiles and multiple broader profiles with $\sigma$ $\sim$ 70 -
80 km s$^{-1}$, indicating the complex nature of the nebula kinematics. A comparison with previously obtained VLT/FLAMES spectroscopy demonstrates that the data agree well in the regions of overlap but the Fabry-Perot data are superior in spatial coverage. A preliminary analysis of the observations finds a new expanding bubble South of R136, with a projected radius of r=5.6 pc and an expansion velocity of 29 $\pm$ 4 km s$^{-1}$.
In addition, the first-time detailed kinematic maps derived here for several complexes and filaments of 30 Doradus allow identification of kinematically independent structures. These data exemplify the power of the combination of a high order Fabry-Perot with a wide-field imager (3 $\times$ 3
arcmin$^2$ GLAO-corrected field of view) for high-resolution spatial and spectral studies. In particular, SAM-FP data cubes are highly advantageous over multi-fiber or long-slit data sets for nebula structure studies and to search for small-scale bubbles, given their greatly improved spatial coverage.  For reference, this paper also presents two appendices with  detailed descriptions of the usage of Fabry-Perot devices, including formulas and explanations for understanding Fabry-Perot observations.

\end{abstract}

\section{Introduction}
\label{Introduction}

This paper introduces a new restricted-use instrument\footnote{
Restricted-use instruments at SOAR (Southern Astrophysical Research
Telescope, a 4.1-meter aperture telescope located on Cerro Pach\'on,
Chile) are those brought to the telescope and owned and maintained
by individual scientists and/or their research groups.  Scientists
outside the team are welcome to use this mode and, if interested,
should contact the first author.} 
for the SOAR telescope, or more precisely,
it describes a new mode inside an existing facility instrument. It
consists of a Fabry-Perot device mounted inside the SOAR Adaptive
Module (SAM) with the SAM imager (SAMI, the existing imager, regularly
used with SAM), to deliver high spatial resolution data cubes, in the optical, with a field of
view of 3 x 3 arcmin$^2$. Presently, there is one Fabry-Perot instrument available for regular
use with SAM-FP, which yields $R \simeq 11200 $ at H$\alpha$. 
This new mode is hereafter referred to
as SAM-FP. 

In \citet{mdo13} a number of recent Fabry-Perot spectrometers
available for astronomy were described, but all of them were
seeing-limited with a spatial sampling that depended on the detector
pixel scale. Fabry-Perot instruments have been used for a range of
sciences but the study of the internal kinematics of emission-line galaxies has been
a major field of interest (e.g. Torres-Flores et al. 2013b, 2014; Alfaro-Cuello et al. 2015). 
 Here, we highlight the potential for various studies,
of the SAM-FP instrument, with a pixel size of 0.045 arcsec/pixel
and ground layer adaptive optics (GLAO) correction down to 0.3 arcsec.
A high-order, high-resolution Fabry-Perot with GLAO can be a unique tool in the
study of kinematics of emission-line objects such as star forming
galaxies, gas flows, planetary nebulae, HH objects, giant H{\sc ii}
regions, among others.  A tunable filter with GLAO, on the other
hand, can be used for studying active galactic nuclei and quasars, outflows, massive
young stars, mass loss processes and Ly$\alpha$ detection studies,
to give a few examples.  In this paper we focus on data taken with
a high-order, high-resolution Fabry-Perot device. In a future
paper, data taken with SAM-FP and a low-order tunable filter will 
be described.

We present Fabry-Perot data cubes for the large-diameter expanding
structures around a giant H{\sc ii} region, the centre of 30 Doradus 
starburst (or 30 Doradus giant star forming region)
also called the Tarantula Nebula,
in the Large Magellanic Cloud.
The study of shells and bubbles, due to stellar winds and supernova
explosions, within giant H{\sc ii} regions is crucial for understanding
the processes of mass and energy transfer from the stellar component to the
interstellar medium and the processes of triggering of star formation. 
Although kinematic studies
of super-giant shells and super-bubbles in giant H{\sc ii} regions are
common in the literature, carried out both with long slit and more recently
with 3D spectroscopy, they have not had the necessary coverage and spatial
resolution to answer a few key questions. Even the origin of the
supersonic motions present in these regions is still a matter of
debate \citep{melnick99}.  Moreover, it is still not understood
if the kinematics of giant H{\sc ii} regions are due to an underlying
diffuse component or to superposed multiple kinematic components.
Having high spatial and spectral resolution
velocity maps, as those delivered by SAM-FP, is crucial to tackle
these problems.

A few authors have studied the kinematics of the ionized gas in the
30 Doradus nebula, e.g. \citet{chu94}, \citet{melnick99},
\citet{redman03}, \citet{torresflores13} and references therein.
All of these studies reported broad and possibly multiple components
widespread in the nebula, which indicate a complex structure.
These studies were either performed (i) with long-slit echelle
spectroscopy (e.g. Chu and Kennicutt 1994;  Melnick et al. 1999),
probing motions and physical conditions in small (pc) and larger
(tens of pcs) scales but covering only a limited number of lines
of sight or (ii) with fibers, as the work of \citet{torresflores13},
with a somewhat larger field of view,  but probing the nebula in
a sparse manner (every 5x5 pc$^2$, with a filling factor of 0.0028).
The advantage of the
present study is that it combines 3D spectroscopy information with
{\it continuous} coverage of the field over a fairly large field of view, with a
pixel size (binned 4 x 4) of 0.18 x 0.18 arcsec$^2$ (or 0.045 x 0.045 pc$^2$). These data show that the kinematic
profiles change on scales of a tenth of a parsec over the
whole field.  The wealth of details seen in Hubble Space Telescope
(HST) images of nearby giant H{\sc ii} regions can now be investigated
using similar quality kinematic data taken with SAM-FP with a laser
guide star.

The main purpose of this paper is to highlight the potential of
Fabry-Perot maps obtained with adaptive optics in the study of 30
Doradus and similar giant H{\sc ii} regions.  A high-order Fabry-Perot
was used to observe the ionized gas within 30 Doradus, in its strongest
optical line: H$\alpha$.  In section \ref{Description}, we first describe
the new mode of SAM with a Fabry Perot, SAM-FP. 
We also include in this section a general description of the data
reduction.  The observations are presented in section \ref{Observations}.
Section \ref{DataAnalysis} compares the present observations with
earlier ones taken with a different instrument.
We describe in section \ref{Results} our results
which include monochromatic, radial velocity, velocity dispersion
and channel maps of the region. We identify a new expanding bubble
in the centre of 30 Doradus and we describe the overall kinematic
features observed, thanks to the high spatial resolution and
continuous spatial coverage of the data.  Section \ref{summary}
summarises our main results. Appendix \ref{What is needed for carrying out Fabry-Perot observations} details all steps necessary for taking Fabry-Perot observations.  Appendix \ref{What is needed for understanding Fabry-Perot observations} provides further information on section \ref{Dataacquisition} and Appendix \ref{What is needed for carrying out Fabry-Perot observations}.  The Fabry-Perot guide for observers given in the appendices is essential because s
uch information cannot be found in one place anywhere else, and this often forms 
a barrier to using Fabry-Perot instruments. The main aims of this paper are, therefore, to describe the 
SAM-FP instrument, to present example observations taken with this instrument 
and finally to provide recommendations on how to best
perform Fabry-Perot observations, with any generic FP instrument.

At the adopted distance of 50 kpc to 30 Doradus, 4 arcsecs corresponds to 1 pc.

\section{Description of the instrument}
\label{Description}

\subsection {SAM-FP: a new mode for SAM}
\label{anewmode}

The SAM instrument \citep{tokovinin08} is a unique
facility instrument mounted on the 4.1 m SOAR telescope at Cerro
Pach\'on. SAM improves image quality by partial correction of
turbulence near the ground using ground-layer adaptive optics (Tokovinin et al. 2010, Tokovinin et al. 2012). This instrument
can feed corrected images either to a visitor instrument or to the
internal wide-field optical imager, SAMI \citep{fraga13}.
SAMI works with a single e2v CCD with 4096 $\times$ 4112 pixel$^2$.
Each pixel is 15 $\times$ 15 $\mu$m$^2$ square, the pixel scale is 45.4 mas and
the total field of view is 3 $\times$ 3 arcmin$^2$.  The CCD is operated with
the SDSU-III controller, which reads the full unbinned chip in 10 seconds
with a noise of 3.8 electrons (without patterns) and a gain of
2.1 electrons/ADU. SAM can provide very sharp images with image quality
as good as $\sim$ 0.3 arcsec in the r-band, in favourable conditions,
after laser corrections. SAMI is regularly mounted on SAM
for use of the SOAR community and has been used with 
success (e.g. \cite{fraga13}). 

SAM was built with space in the collimated beam for the installation
of an etalon, to enable future Fabry-Perot observations, which had
not been implemented until now.  Presently, there is one Fabry-Perot
instrument fully tested for use in SAM, with a pupil size of
65 mm (ICOS ET-65\footnote{This was fabricated in 1988 by
Queensgate, nowadays called IC optical Systems (ICOS), for the Anglo
Australian Observatory and it is, presently, in an extended loan
to the University of S\~ao Paulo, kindly made possible by the present
director Dr. Warrick Couch and the previous director Dr.  Matthew
Colless.}).  It has a mean gap of 200 $\mu m$
and, at H$\alpha$ 6562.78 \AA,  an interference order $ p\simeq 609$. For the 
observational run described in this paper, an effective
finesse $F \simeq 18.5$ and a resolution $ R \simeq 11200$ (at H$\alpha$) were measured; 
however these values may change slightly depending on the accuracy of 
the Fabry Perot plates' parallelism.
The Fabry-Perot plates have a broadband coating covering the range 400-700
nm with a reflectivity of 90\%. This coating was inspected in 2015
by ICOS and was found to be in good condition. 

The etalon is controlled by a CS100 controller (Serial Number
8030), borrowed from the Anglo Australian Observatory.  
The controller is attached to the SOAR's Nasmith cage by a mechanical adaptor specifically built for it, 
placed in such a way that one can access and adjust it while simultaneously inspecting
the etalon. 
The computer that runs the CS100 is located on one side
of the Nasmith platform where SAM is permanently mounted.  The CS100
contains two channels, X and Y, to control the parallelism between the
Fabry-Perot plates, and a Z channel to control the spacing.  A
similar CS100 to the one used here is described in detail by 
\citet{vei10}.
 
SAM's original concept already included a mechanism to put the Fabry-Perot in and
out of the beam, which can now be used when doing Fabry-Perot observations.
It was necessary to design and fabricate a mechanical plate adaptor in order to mount the Fabry-Perot device
inside SAM. This adaptor can be removed at any time but it is now
regularly placed into SAM. 
It was devised in a way to allow the Fabry-Perot to be easily placed
in and out of the instrument, facilitating the task of checking the
parallelism while tuning the CS100, which is located immediately
beside it.

\subsection{Data Acquisition}
\label{Dataacquisition}

The existing control software of SAMI was modified (from a module
previously written for the Brazilian Tunable Filter Imager, Mendes
de Oliveira et al. 2013) in order to create a script for taking a
scanning sequence. 
The script was used during the setup of the
Fabry-Perot, for on-sky observations, as well as for acquisition
of spectral calibrations.  The reasons for each command and
descriptions of how the Fabry-Perot parameters are actually determined
are described in the appendices. 
In summary, the script allows computing
the exact value for the free spectral range of the Fabry-Perot
device, 
finds the ideal number of steps needed to scan the free spectral range and defines a scanning
sequence.  
As for a grism, the wavelength range and
the spectral resolution are the important parameters that define
a Fabry-Perot device.  However, for a Fabry-Perot,
the effective spectral range is in fact the free spectral range
(FSR), see equation (\ref{lfsr}) below, and the spectral resolution
depends on the interference order p and the effective finesse F, which
follows relation (\ref{spectralresolution}) below.

\subsubsection{Determining the free spectral range, the effective finesse and spectral resolution}
\label{computingparameters}

In order to compute the FSR at a given wavelength $\lambda$, the
interference order p should be known and it is linked to the other
quantities by the basic Fabry-Perot equation:

\begin{equation} 
 \rm{p=\frac{2ne \cos \theta}{\lambda}}
\label{necosipl} 
\end{equation}
 
where $ \rm{n (T=0^oC, P=1\  atm)\simeq\ 1.0003}$ is the index of the
air layer between the plates of the Fabry-Perot device, $\rm{e}$ the
inner separation between the coated plates, $\theta$ the incidence angle
and $\lambda$ the wavelength.   This is valid 
in the case where  the so called additional phase-lag $\psi(\lambda)$, defined in the appendix, can be neglected; see section
\ref{Wavelength calibrations and additional phase-shift effect} and equation \ref{ppsy} for the general case.
From equation (\ref{necosipl}), it follows
that for a given n, $\rm{e}$ and $\theta$,
the FSR at the interference order p
and at the wavelength $\lambda$ can be computed by:
\begin{equation}
{\rm \Delta\lambda_{FSR}}  =
 \rm{\frac{\lambda}{p}\left(\frac{1}{1-({1}/{p^2})}\right) }
\label{lfsr}
\end{equation} 
Conversely, for a given n, $\lambda$ and $\theta$, again from equation
(\ref{necosipl}),  the FSR at the interference order p and
at the inner separation of the plates $\rm{e}$ can be computed by:
\begin{equation} 
 \rm{\Delta e_{FSR} =\frac{e}{p} = \frac{\lambda}{2 n \cos \theta} }
\label{efsr}
\end{equation} 
$ \rm{\Delta\lambda_{FSR}}$ is then the FSR in \AA\ (if $\lambda$ is
given in \AA), which can be easily related to the FSR in
velocity units (see numerical values below), while $ \rm{\Delta e_{FSR}}$ is the
increment of plate separation necessary to scan the FSR.

As described in the previous section, one Fabry-Perot device is
presently available, ICOS ET-65, with a high interference order 
p $\simeq$ 609, at H$\alpha$, allowing coverage over a FSR $\simeq$ 
10.8 \AA\ (following equation \ref{lfsr}), which corresponds to $\simeq 492$
km s$^{-1}$. 

In order to select only one or a few interference orders, an
interference filter selecting a passband broader than the FSR has
to be placed in the optical path, generally at (or close to) the
focal plane to avoid wavelength dependence.  Ideally the central
wavelength of the interference filter should correspond to the mean
velocity of the source.  For instance, for the observations of 30 Dor
 with a velocity restframe of $V_s$=267 km s$^{-1}$ (Torres-Flores et 
al. 2013, placing H$\alpha$ at $\lambda$=6568.65 \AA),
we selected an
interference filter centred at $\lambda_c$=6568 \AA\ having a
passband of 19 \AA\ (which is, in the case of the  
Fabry Perot used here, almost twice the FSR).

The resolving power, also called spectral resolution $ \rm{R_{\lambda}}$ (or simply R) of the Fabry-Perot is provided by the following equation:
\begin{equation}
 \rm{R_{\lambda} = p F}
\label{spectralresolution}
\end{equation}
We thus need to know the interference order p and the effective finesse
F.  In practice, both parameters are roughly known and they are
given by the manufacturer. The effective finesse needs nevertheless
to be measured with accuracy.  As explained in the end of Section \ref{Measuring the Finesse}, the
effective finesse can be simply obtained dividing the FSR by the width of an
arc line (e.g. the Ne I emission line 6598.95 \AA) in a spectrum extracted
from the calibration cube.  The spectral
resolution can be inferred from the width of a calibration arc line. 
The high order Fabry-Perot for use with
SAM-FP has $ \rm{p \simeq 609} $, at H$\alpha$ 6562.78 \AA. During the observational run when data
for this paper was obtained, the effective finesse of the FP was measured to be  
$ \rm{F \simeq 18.5}$, providing a spectral resolution
of $ \rm{R \simeq 11200}$.
Further information on the nature and on the determination of the effective finesse is provided in section \ref{Measuring the Finesse}.

\subsubsection {Taking calibrations and on-sky data}
\label{takingcalibrations}

Once we know the FSR and effective finesse, then the optimal number of steps
to scan a FSR has to be computed and the scanning sequence has to
be defined.
For the
observations of 30 Doradus described in this paper, we decided to
slightly oversample by 5-10\%, choosing a scanning sequence of 40 steps
with a scanning step of $\sim$ 12.8 km s$^{-1}$.  The same number
of channels must be used for on-sky and calibration scanning
sequences, except for the one calibration cube that is obtained for
determination of the FSR and effective finesse, which should be highly
oversampled and should cover more than one FSR (see Appendix \ref{What is needed for understanding Fabry-Perot observations}). 

Spectra of calibration arc lamps (those from the SOAR calibration
unit) need to be taken in day time, ahead of the first observing
night, for wavelength calibration and for computation of the several
Fabry-Perot parameters that will define the observation strategy.
Suitable interference filters should be illuminated by the calibration
lamps, so that the final spectra will contain at least one arc line
with similar wavelength to that of the observed emission-line
(H$\alpha$ in our case, redshifted due to the velocity of the
target). In our case, we chose to use the Ne I 6598.95 \AA\ emission
line for calibration and the best option was then to use the
interference filter BTFI 6600.5/19.3, available from the BTFI filter
set.

A reference wavelength calibration has to be repeated at the position
of the object on the sky ideally at the start and at the end of
each observation to account for possible mechanical flexures of the
instrument as well as changes in the air index inside the FP plates
due to temperature, pressure and humidity changes along the night.
In section \ref{Datareduction} details are given about how to 
perform wavelength calibration of the data.

We described above and in the Appendices
how to compute the parameters
necessary for setting up the Fabry-Perot and taking data.
In practice, a Python program was written in order to define the 
acquisition sequence for a Fabry Perot data cube. It produces a shell script that has to be run each 
time the observer wants to take a on-sky observation or a calibration. 
The Python program is available in the SAMI machine and upon request.
It is self explanatory and easy to use.  The input parameters to 
the script are given in Appendix \ref{What is needed for carrying out Fabry-Perot observations}.

We should note that a feature was included in the control
software of SAMI in order to allow pausing/stopping while an exposure
is being taken.  This was necessary, given that there are laser
interruptions that are mandatory during a given night. In addition,
the observation can be paused/stopped due to passing clouds. When
the stop or pause button is pressed, the ongoing exposure is then
finished and read out before the action (pause or stop) takes place.

A mode allowing 8 $\times$ 8 binning was needed for testing, at the
commissioning, and this was then included in the software. A 4
$\times$ 4 binning mode (an option that already existed previously
in SAMI but had not been implemented) was included in the new control
software and has been used for the Fabry-Perot observations presented
in this paper. Indeed, a binned pixel of size 0.18 arcsec still allows a
good sampling of the best seeing and considerably increases the signal-to-
noise ratio (SNR)
per unit surface.

\begin{table*}
        \centering
        \caption{Journal of Perot-Fabry observations}
        \begin{tabular}{lrr} 
                \hline
                Observations        & Telescope  & SOAR \\
                    & Instrument &   Fabry-Perot inside SAM \\   
                    & UT Date       & March 18$^{th}$ 2015 \\   
                    & FWHM of stars (laser corrected)    & $\sim$0.5-0.7" \\
                    & Mean scanning lambda of 30 Doradus obs & 6568.78 \AA \\
                    & 30 Doradus systemic velocity & 267 km s$^{-1}$ \\
Interference Filter & Central Wavelength & 6568 \AA  \\
                    & FWHM               & 19~\AA   \\
                    & Transmission & 0.70  \\
Calibration         & Ne I reference line &  6598.95~\AA \\
                    & Central lambda of filter used to select Ne line & 6600.5~\AA \\
                    & Width of filter used to select Ne line &   19.3~\AA \\
Fabry-Perot         & Company and ID   &  ICOS ET-65 \\ 
                    & Interference Order &609 @ 6562.78 \\
                    & Free spectral range at H$\alpha$  & 10.8~\AA (492 km s$^{-1}$) \\
                    & Effective finesse & 18.5 at 6598.95~\AA\tnote{c} $^c$ \\
                    & Spectral resolution at H$\alpha$ & 11200\tnote{c} $^c$ \\
Sampling            & Number of Scanning Steps & 40\tnote{c} $^c$ \\
                    & Sampling Step   & 0.28~\AA (12.8 km s$^{-1}$)\tnote{c} $^c$ \\
Detector            & e2v & CCD 4096 $\times$ 4112 \\
                   & Total Field of View & 120" $\times$ 120" \tnote{a} $^a$ \\
                  & Pixel size & 0.18"  \tnote{b} $^b$\\
                    & Readout noise (unbinned) & 3.8 electrons \\
                    & Gain & 2.1 electrons/ADU \\ 
Exposures times     & Total on-target exposure & 80 minutes \\
                    & Total exposure time per channel & 120 s \\
                \hline
        \end{tabular}
        \begin{tablenotes}
                \small
                \item[a] $^a$ The field of view is limited to 2 $\times$ 2 arcmin$^2$ by the interference filter used.
		\item[b] $^b$ The pixel size of SAMI is 0.045 arcsec. Pixels were binned 4x4.
		\item[c] $^c$ These values may change slightly from one observational run to the next, given that they depend on the 
		                         accuracy of the FP plates' parallelism (see appendix) and this has to be checked every time the instrument is used. 
        \end{tablenotes}
        \label{table1}
\end{table*}

\subsection {Data Reduction}
\label{Datareduction}

The raw images obtained with SAM-FP are transformed into scientifically
useful data cubes after following several procedures 
described below.  Each FITS file obtained with SAMI has four
extensions (the CCD controller has four amplifiers). Before the
combination of the four fits extensions into one fits image, 
we estimate the
background using the overscan region by fitting a 3rd degree
polynomial along the columns' direction and subtracting it from
each column.  This procedure is followed for all data including
bias, dark, flat, calibration and science files. After combining
the extensions we then subtract the bias from the science frames
and correct them by the normalised flats.  It is important to note
that SAMI's CCD has a strong instrumental feature that is not always
present but can greatly affect the images. These are described as
``arcs" in internal SAMI documents. They can be removed from the
science images by subtracting suitably scaled dark images from the
science frames, before flat fielding.  It is, therefore, highly
recommended that at least 10-20 darks are taken per night, with
exposure times of 3-10 minutes, for this purpose.  This feature was
not seen in the data presented here, so no correction was needed.
For science images with exposure times longer than several seconds,
it is important to do cosmic ray cleaning. In our case, all procedures
described above were performed using IRAF\footnote{IRAF is distributed by
the National Optical Astronomy Observatories, which are operated
by the Association of Universities for Research in Astronomy, Inc.,
under cooperative agreement with the National Science Foundation.},
except for the procedure of combination of the fits extensions,
performed by a program kindly provided by Andrei Tokovinin.

The final calibration and science images were then converted into
data cubes. Having the
calibration cube (in our case the Ne I data cube) in hand, 
the phase correction was performed with the
main goal of re-arranging the spectral information in the science
data cube. As it is well known, the spectral information in each
pixel, for Fabry-Perot data, is radially shifted with respect to the next,
as one moves in the cube, along the spatial direction.  This shift follows a
parabolic surface centred on the Fabry-Perot optical center. 
In addition, because Fabry-Perot data are periodic, the
spectrum repeats itself.  Three steps are required for the phase
correction: phase-map extraction, fitting and application.  These
steps have already been described previously in the literature,
e.g. \citet{Atherton82}, \citet{Bland87}, \citet{Amram89}.
After phase correction we then have a data cube where each frame represents
one wavelength (modulo the FSR, given the cyclic nature of the 
Fabry-Perot device).  
  
For wavelength calibration we use two calibration cubes taken at
different wavelengths. In fact, only one cube would be necessary for
calibration of data taken with a high-order Fabry Perot 
but a second one can be useful for interference order checking.
In our case we use a Ne Lamp with two filters that contained the
following lines respectively, 6598.95 \AA\ (BTFI filter 6600.5/19.3)
and 6717.04 \AA\ (BTFI filter 6745.5/38.6),  but the latter is only for
order checking.  For that, the calibration cube (in
our case the 6598.95 \AA\ data cube) has to cover more than one free
spectral range so that the line will appear in two different orders
and the second cube (in our case the 6717.04 \AA\ data cube) can
cover one free spectral range only.  We also need to extract a 1D
spectrum from these cubes at the centre of the rings.  Section 5.9
of \citet{Atherton82} describes in detail the procedure for wavelength
calibration of tunable filter images that has been adapted for our
use and built into a script. 
After wavelength calibration the cube is then ready for night-sky line subtraction and derivation
of the monochromatic, velocity and dispersion maps.  

All command line scripts needed to perform the tasks
described above are available at the SOAR observatory, with supporting 
documentation.

\begin{figure}
 \centering
 \includegraphics[width=0.47\textwidth]{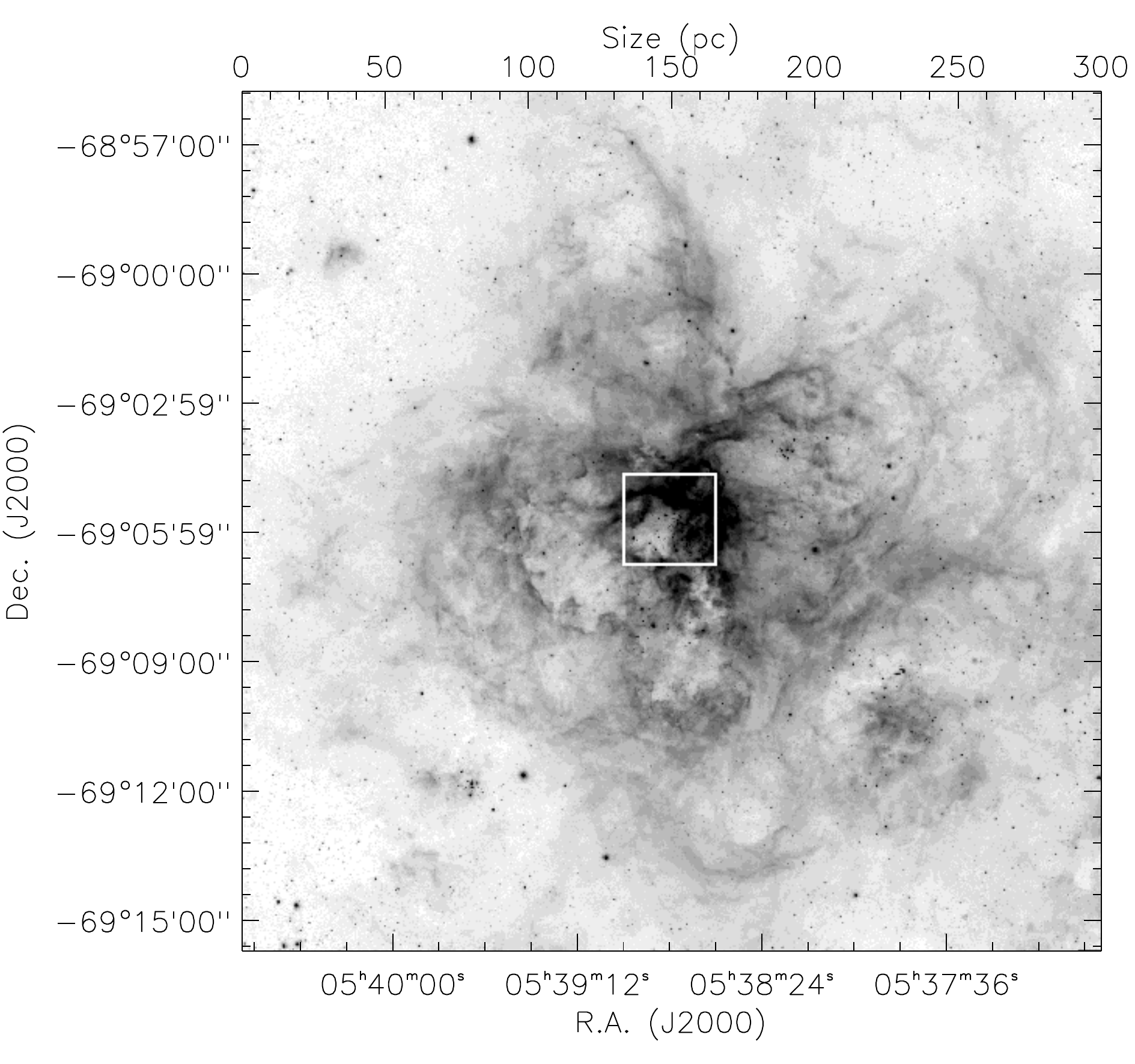}
\caption{
H$\alpha$ image taken at the T80-South telescope during comissioning of its wide-field camera, with images with a field of view of 1.4 $\times$ 1.4 degree$^2$. The large image above is a subset of it, covering a $\sim$20  $\times$ 20 arcmin$^2$ field centred on R136 and the white box indicates the central 2 $\times$ 2 arcmin$^2$ of the image observed with SAM-FP.
}
 \label{field}
\end{figure}

\section {Observations of 30 Doradus}
\label{Observations}

The observations of 30 Doradus, centered at 
RA 05 38 47.4 and Declination --69 05 43, were taken on UT date March 18th, 2015, with the
high-resolution scanning Fabry-Perot interferometer described above,
placed inside SAM, fed to SAMI.  Although the total field of view
of the instrument is 3 $\times$ 3 arcmin$^2$, due to the smaller size of the H$
\alpha$ interference filter, the useful field of view was 2 $\times$ 2 arcmin$^2$.
We used a 2 $\times$ 2 inch$^2$ interference filter from the Cerro Tololo list,
centred at 6568 \AA\ (filter 6568-20), with a FWHM of 19 \AA\ and with transmission of
$\sim$70\% (according to the list of CTIO filters).  
The free spectral range of the interferometer (492 km s$^{-1}$)
was covered in 40 scanning steps. The setup and observing details are
summarized in Table \ref{table1}.

The observations were taken under photometric conditions.  The
seeing during the observations varied between 0.7 and 0.8 arcsec.
GLAO correction lead to an improvement of spatial resolution
and flux of up to $\simeq$ 2, yielding a FWHM of 0.5-0.7
arcsec.  The spatial resolution of the cubes was then determined
by the laser corrected FWHM with a mean value of 0.6 arcsec or 0.15 pc.
The instrumental profile, as measured from Ne calibration lamp lines
was 0.586\AA\ or 26.8 km s$^{-1}$, which corresponds to a spectral
resolution of about 11200 at H$\alpha$.  The last four channels
were taken after a laser interruption of a few minutes and after
some technical problems which took about 45 minutes to be fixed.
The data for the last channels were taken under higher
airmass ($>$ 1.5) and slightly worse seeing (0.7 arcsec after correction).
This, however, represented only 10\% of the data and it did not
degrade significantly the quality of the cube as a whole.

Fabry-Perot observations are usually taken with short exposure times
and in a number of sweeps, to account for
transparency and seeing variations. However, given that we are
observing with a classical CCD (not an EMCCD nor a photon-counting camera),
we chose to make just one sweep of
40 channels and with fairly long exposure times of 120 seconds each, to
minimize the readout noise (see also section \ref{How to choose the observing time ?}). The night sky lines have been identified by plotting a histogram of wavelengths and picking the most frequent values (given that the sky lines are present in every pixel). Knowing their wavelengths and intensities, night sky lines were then subtracted. Alternatively, a nebulae-free data cube beside the object (new pointing) could have been observed to directly subtract the night sky lines (see section \ref{How to get rid of the night sky line emission ?}) but this was not done in the case of these 30 Doradus observations.

Figure \ref{field} shows the region observed
in the present study.  The background image is an H$\alpha$ observation
taken during commissioning of the robotic telescope
T80-South\footnote{T80-South is an 80-cm telescope mounted at Cerro
Tololo, which has a CCD camera with a 9.2k $\times$ 9.2k pixel$^2$
CCD yielding a 1.4 $\times$ 1.4 degree$^2$ field of view (with
0.55-arcsec pixels). The telescope was built by the companies AMOS
and ASTELCO and the CCD camera by Spectral Instruments.}.  A cut of
20 $\times$ 20 arcmin$^2$ centred on 30 Doradus was extracted from the
T80-South image and it is shown in Figure \ref{field}.  The area
of 2 $\times$ 2 arcmin$^2$, observed with the Fabry-Perot, is
indicated with a white box.  The bright star cluster R136 is included
in the area observed.

\section {Data Analysis}
\label{DataAnalysis}

\subsection {Comparing Fabry-Perot data with previously published H$\alpha$
kinematic data of 30 Doradus}

\begin{figure}
\includegraphics[width=0.47\textwidth]{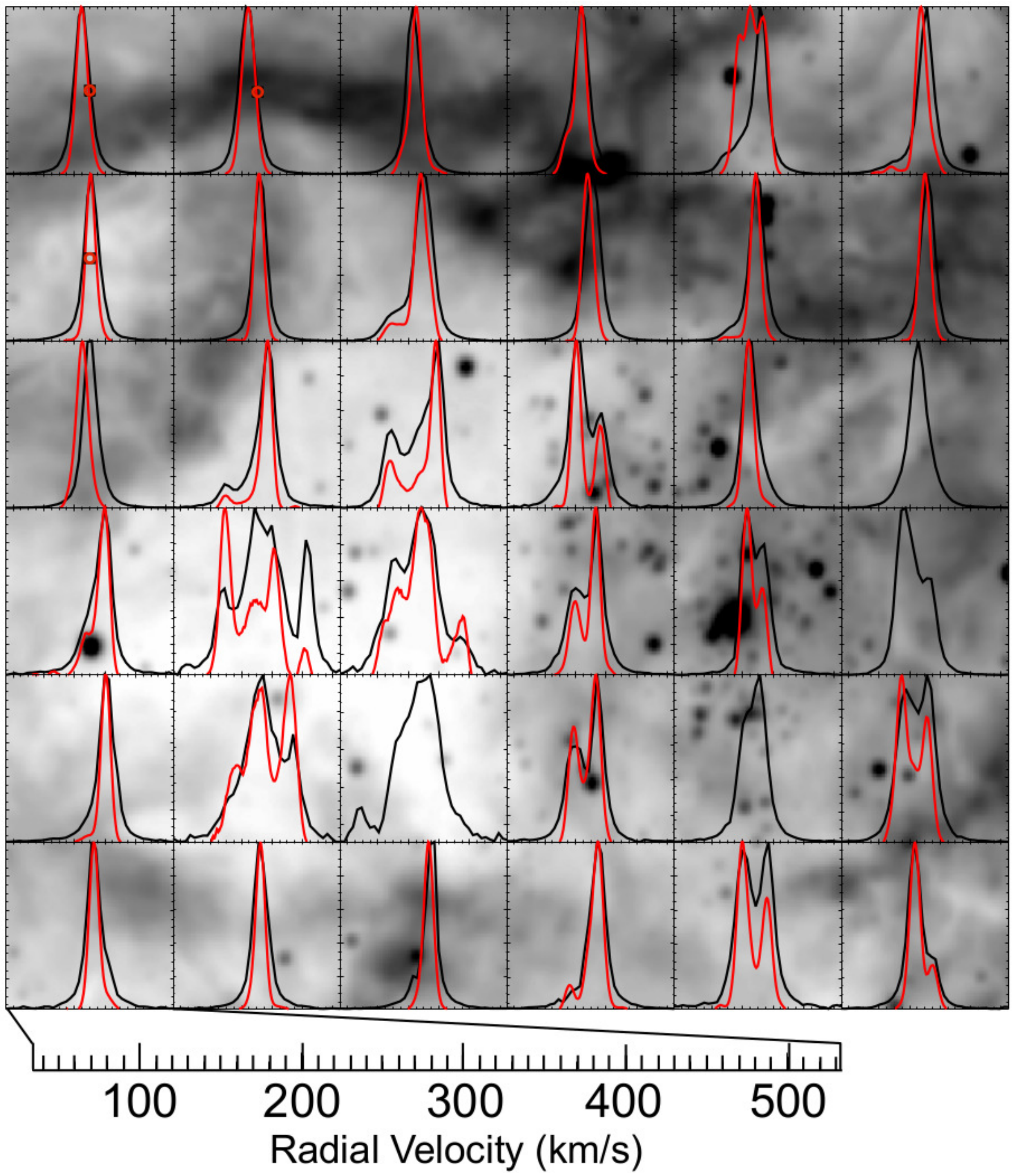}
\includegraphics[width=0.47\textwidth]{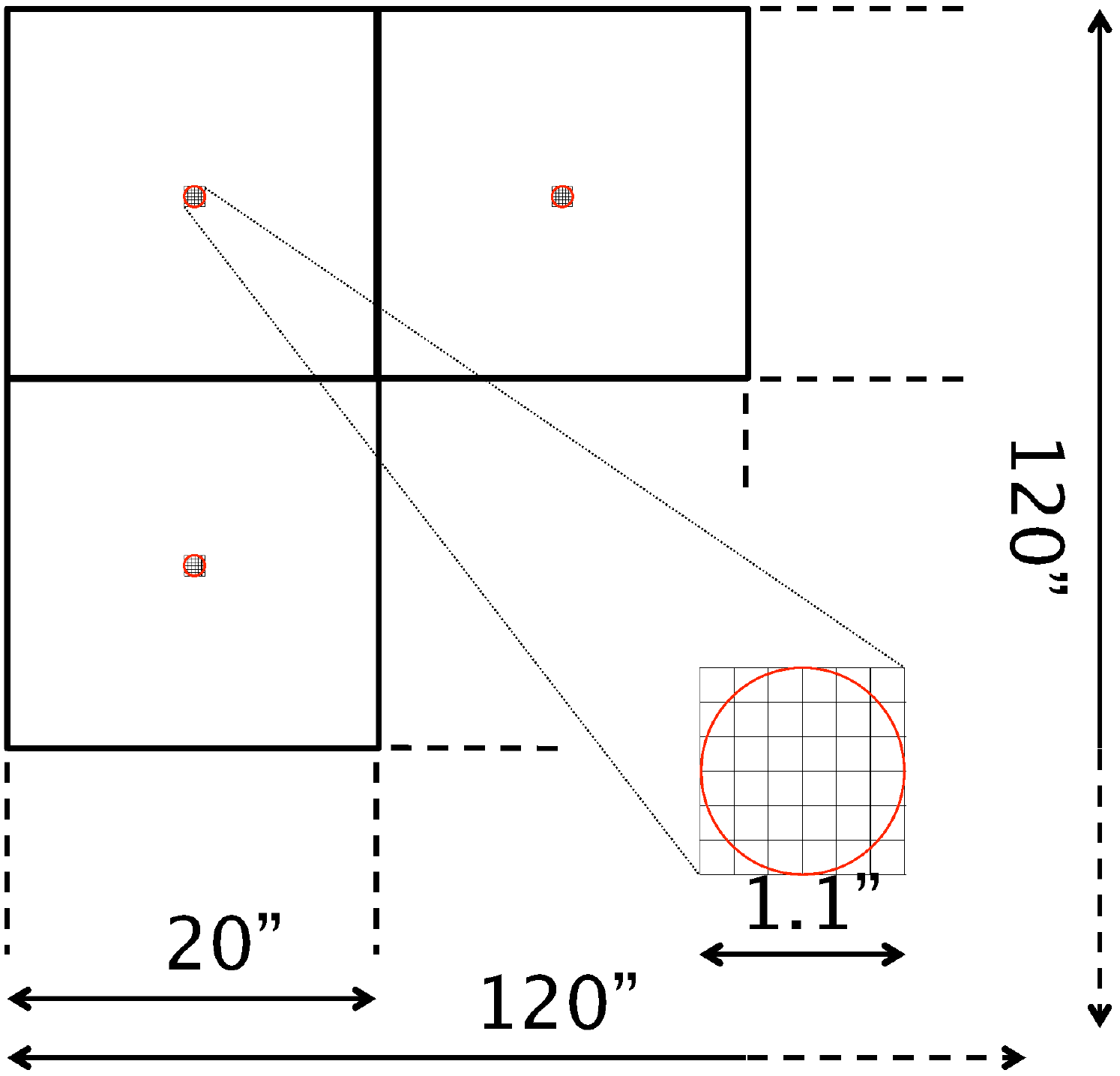}
\caption{Upper panel: central region of 30 Doradus. The red profiles correspond to the
H$\alpha$ emission-line VLT data, extracted from fig. 3 of
\citet{torresflores13}.  These correspond to the
integrated spectrum within a fibre of 1.2 arcsec diameter, placed
at the centre of each 20 $\times$ 20 arcsec$^2$ box (the position of the fibre is depicted with
a red circle in the three upper left boxes - also see lower panel for the sketch  
representing the approximate fibre diameter). Boxes with no 
red profiles correspond to broken fibres.
The black profiles correspond to the average profiles of the FP data extracted from a
square aperture of 1.1 $\times$ 1.1 arcsec$^2$ located at the centre of each box, in an 
attempt to mimic the
VLT data (the 1.1 $\times$ 1.1 arcsec$^2$ area within which the profiles were averaged is best seen in the sketch 
of the lower panel). The background image corresponds
to an H$\alpha$ image of 30 Doradus taken at ESO NTT (see text).
The velocity scale is shown in one case, on the lower left, and it is the same for all profiles.
Lower panel: sketch of the fields shown in the upper panel.  }
\label{comparison}
\end{figure}

The Fabry-Perot H$\alpha$ data cube of 30 Doradus was compared with
a data cube obtained with the FLAMES instrument at VLT and the
Giraffe spectrograph in MEDUSA mode, published by \cite{torresflores13}. 
About 900 fibers were used to obtain kinematic information
over a field of view of 10'$\times$10', centred on R136, with
a spectral resolution of $ R \sim 17000 $ (compared to $ R \simeq 11200$ for
the Fabry-Perot data).  The distance between each fiber for the
FLAMES/VLT setup was 20 arcsec.
In the case of the Fabry-Perot data, we observed a
field of view of 2'$\times$2', with a pixel scale of 0.18 arcsec 
and with a median FWHM of 0.6 arcsec.
Therefore, the spectral resolution of the Fabry-Perot data is about
30\% poorer but the spatial coverage or spatial resolution 
(or more precisely, the 
number of independent measurements) 
is more than 30$^2$ times higher
than that for the VLT data (which has a measurement taken with a 
MEDUSA fiber of 1.2 arcsec diameter at every 20 arcsec). 
For this reason, and
in order to do a fair comparison, we have extracted from the Fabry-Perot 
data, squared regions of 1.1''$\times$1.1'' that are 
placed at the same positions
as the MEDUSA fibers were. For each region we have co-added the profiles,
to get a mean H$\alpha$ profile of the region, which can then be
compared with the MEDUSA fiber data.
The red profiles in Figure
\ref{comparison} show the FLAMES/VLT H$\alpha$ data taken
from \cite{torresflores13} and the black profiles represent the Fabry-Perot data.
All profiles have been normalised to the integrated flux of the line in each
specific region or fiber. 
In both cases the background image corresponds to an H$\alpha$
observation of 30 Doradus taken with the ESO NTT/EMMI instrument under program
70.C-0435(A).

Inspecting the profiles shown in Figure \ref{comparison},
narrow- and broad-emission-line profiles can be seen at different
regions of the nebula. The narrow profiles are generally
associated with H$\alpha$ filaments. Broad profiles have been previously reported
by \citet{chu94} among other authors. The broadest profiles can
be seen to the East of R136, and they may be associated with a
cavity that has distinct identifications given by different authors, e.g. it is
called {\it region H} by \citet{melnick99} and {\it Christmas Tree}
by \citet{demarchi14}.  It is a region with an additional
extinction component as compared to the neighbouring regions and it
resembles the silhouette of a Christmas tree.  

Comparing the black and red lines in Figure \ref{comparison} we are able to check if the
Fabry-Perot observations reproduce well the VLT/FLAMES results.
Despite having about 30\% lower spectral resolution (R=11200 against R=17000,  
readily seen by the narrower widths of the VLT profiles), the
Fabry-Perot H$\alpha$ profiles clearly display quite similar shapes to the
previously published data.  For instance, from top left, boxes
(3,2), (3,3) and (2,3) display the same blue component in both
cases.  Also, the shapes of the H$\alpha$ profiles in boxes (4,4),
(4,5) and (5,4) are similar for both datasets.  In addition, the general behaviour
of the emission line profiles in the central region of the {\it Christmas
Tree} is also consistent with results in the literature (e.g. Chu 
and Kennicutt 1994, Melnick et al. 1999).

There are, however, a few notable differences in the line profile shapes between
the VLT and the Fabry Perot data cubes specially in the Southeastern region of the {\it Christmas Tree}, 
e.g. for box (2,4). These differences are real, given that particularly in 
this region the profiles vary strongly in small scales.
The great variation from one frame to the next in the region of the {\it Christmas Tree},
indicating very small scale fragmented velocity features, will be discussed in 
detail in Section \ref{A brief description of the kinematics of the observed region} and in Figure  \ref{profiles}.

In order to do a fair comparison between the SAM-FP and the VLT/FLAMES
data cubes of 30 Doradus, we centred the integrated H$\alpha$ emission
line from the SAM-FP at the exact same radial velocity of the integrated
H$\alpha$ emission line profile of the VLT/FLAMES cube, over the
region of overlap.  The procedure undertaken to match the radial
velocities of both data sets was the following.  First, we derived
a sub-cube of the VLT/FLAMES H$\alpha$ data cube, which covered the
same 2 $\times$ 2 arcmin${^2}$ field of view of the SAM-FP observations
(Figure \ref{comparison}, black lines). We derived the integrated H$\alpha$ profile for
this region,  fitted a single Gaussian to it and obtained a central
wavelength of 6568.72 \AA, which corresponds to 270.7 km s$^{-1}$.
We note that this value is slightly higher than the radial velocity
derived by \citet{torresflores13} of 267.4 km s$^{-1}$, but
that was for the whole VLT data cube (10 $\times$ 10 arcmin${^2}$).  
We then obtained the integrated
H$\alpha$ profile of the SAM-FP data cube, fitted a Gaussian to it
and measured a central wavelength that we matched to 6568.72{\AA}.
Therefore, the zero point of the radial velocity of our dataset is
tied to that derived by \citet{torresflores13}, for the same
region.

Once we have compared both data sets and they show good general
agreement, we proceed to analyse the Fabry-Perot data alone, for which
we have continuous spectroscopic information with a scale of 0.18
arcsec (0.045 pc).

\begin{figure}
\includegraphics[width=\columnwidth]{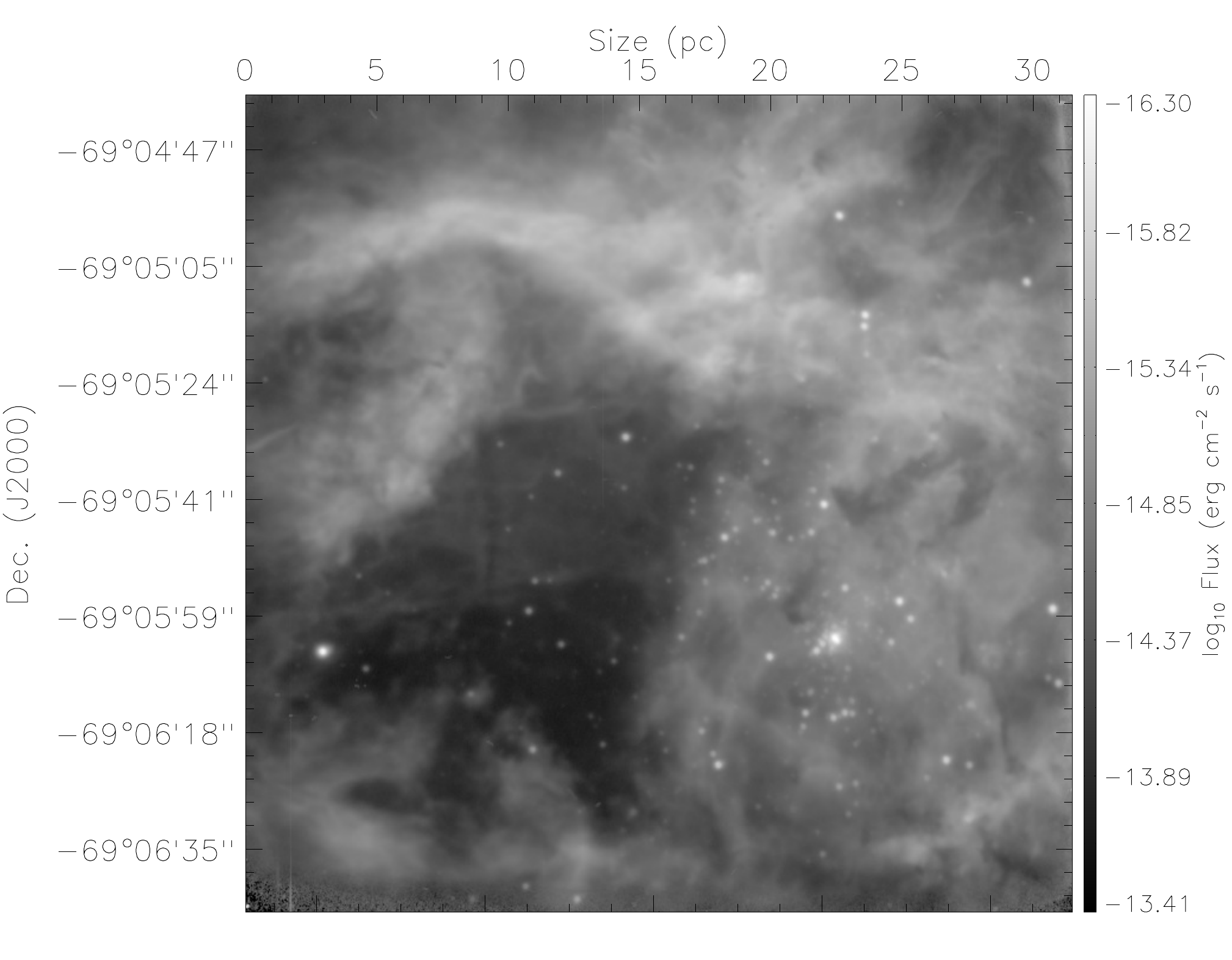}
\includegraphics[width=\columnwidth]{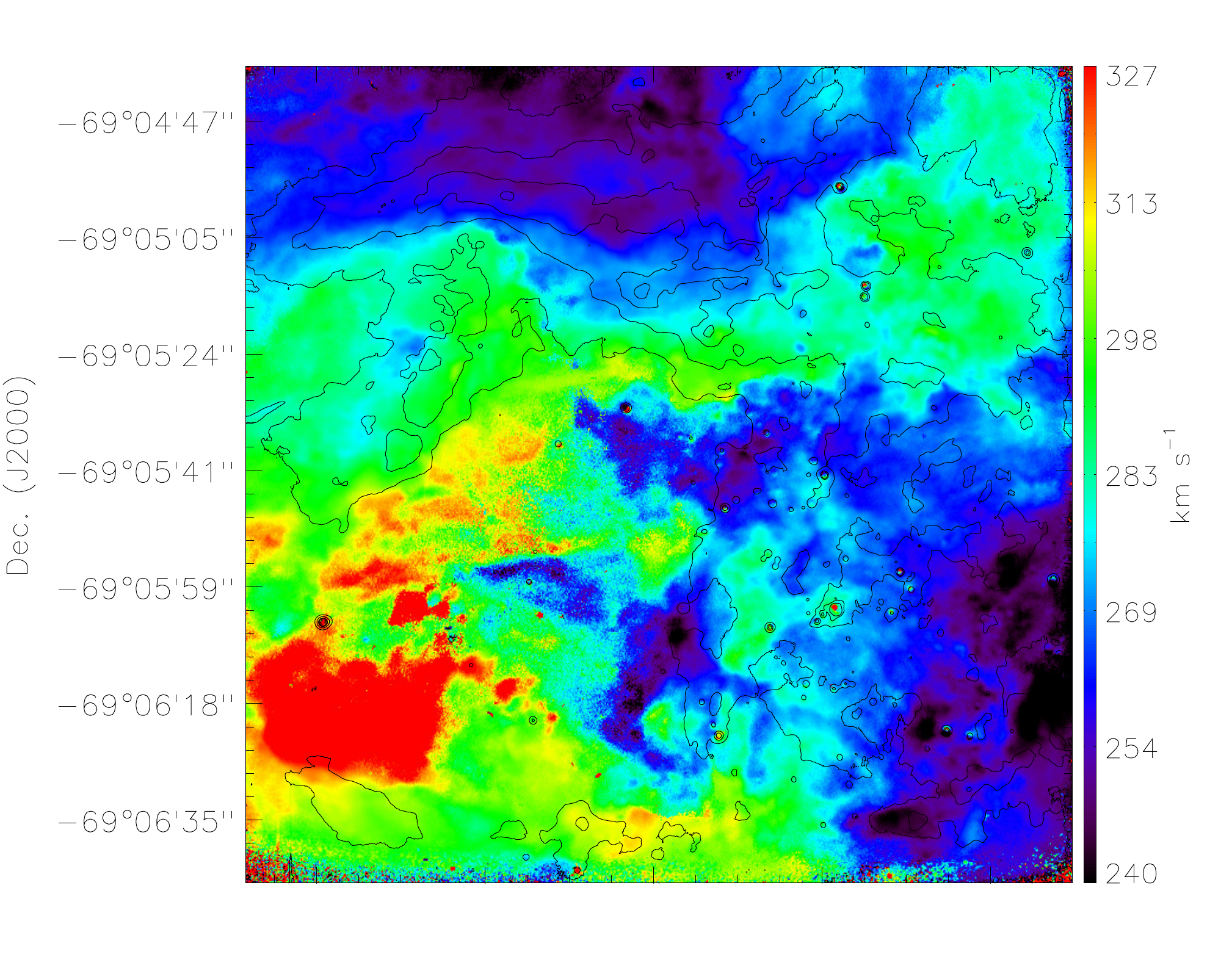}
\includegraphics[width=\columnwidth]{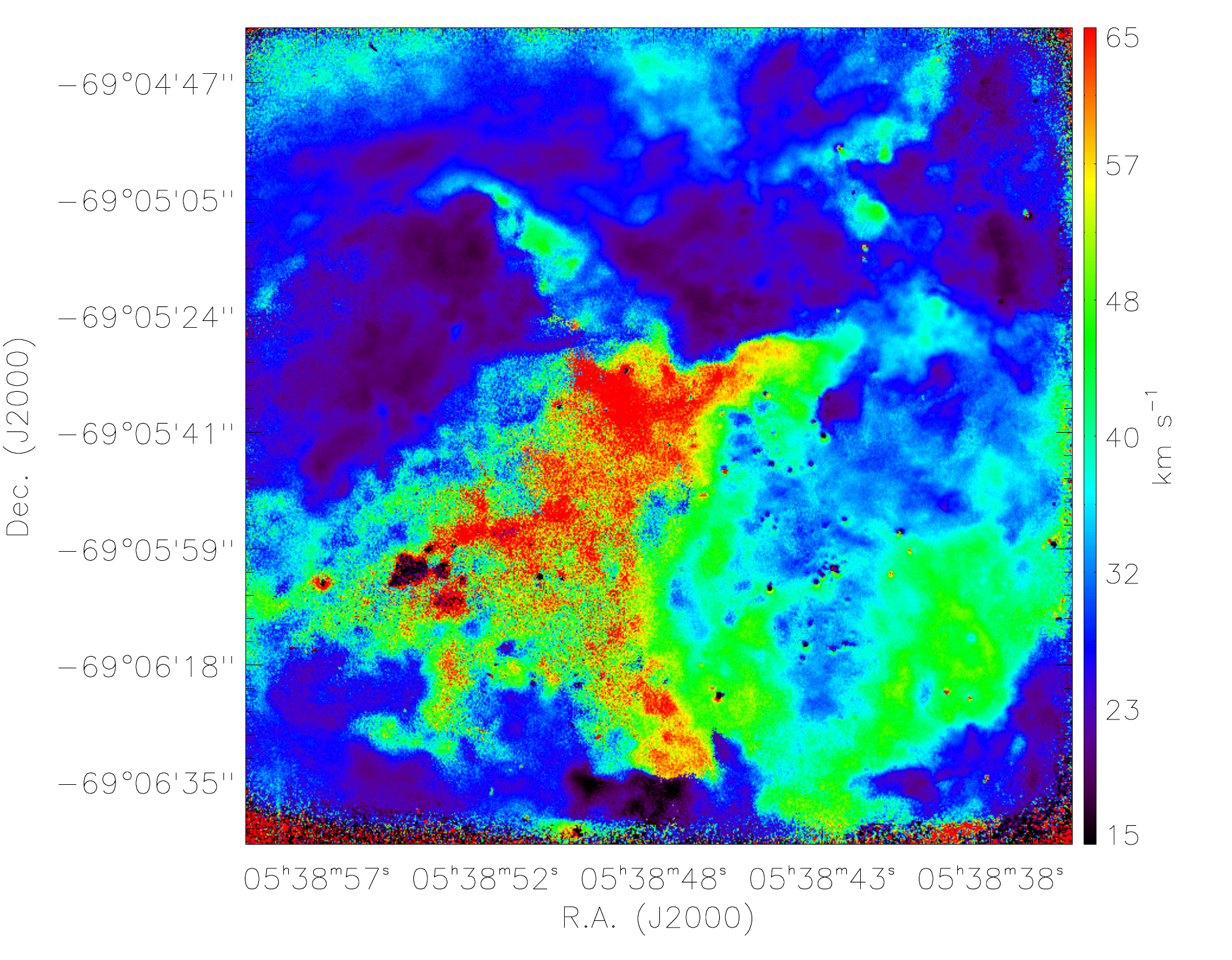}
\caption{Monochromatic image, radial and dispersion velocity maps of 30 Doradus. 
Top: H$\alpha$ monochromatic map derived from the Fabry-Perot data cube, as explained in the text. 
This map is flux
calibrated (see Section \ref{fluxcalibration}). Middle panel: 
Velocity field for the same region as shown in the top panel. The scale is shown in km s$^{-1}$. The contours show the following flux levels (from the monochromatic map): 8 $\times$ 10$^{-16}$, 2 $\times$ 10$^{-15}$, 5 $\times$ 10$^{-15}$, 8 $\times$ 10$^{-15}$ erg s$^{-1}$ cm$^{-2}$. Bottom: 
Velocity dispersion map of the same region. The scale is shown in km s$^{-1}$.
The map was corrected by instrumental and thermal widths.
}
\label{mono_vel_sigma}
\end{figure}

\begin{figure}
\includegraphics[width=\linewidth]{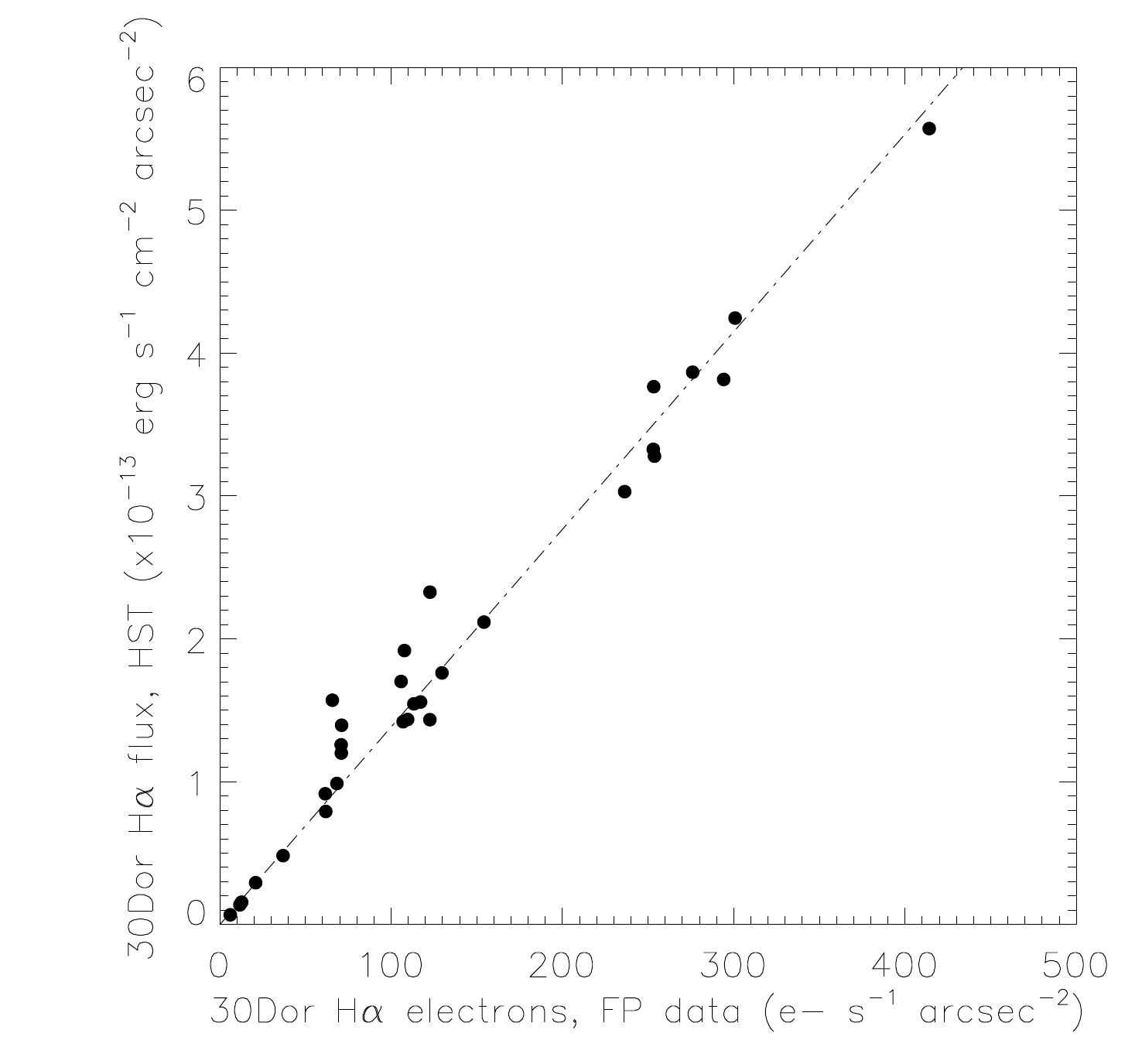}
\caption{Flux calibration of the SAM-FP data of 30 Doradus using
an archival HST image of the region.  In the x-axis
instrumental fluxes of 31 regions located across the FP field,
within an aperture of 3 arcsec in radius, are computed and plotted 
and in the y-axis the
corresponding fluxes for the same regions obtained from the HST
image are given.  }
\label{correlation}
\end{figure}

\subsection{Fitting Gaussians to the H$\alpha$ data cube}
\label{fitting_gaussians}

We have fitted a single Gaussian to each observed H$\alpha$ profile
in 30 Doradus using the code {\sc fluxer}, an interactive routine
in {\sc idl}. This
uses the package {\sc mpfit} to fit a modelled Gaussian to the
observations, which allows us to determine the center, width and
flux of the Gaussian. In this case, the centre can give us information
regarding the radial velocity of the gas and the width can be used
to characterise the overall complexity of the H$\alpha$ profiles,
indicating where there may be a superposition of different kinematic
structures or which regions may be affected by turbulence. In this manner we then computed the
radial velocity and dispersion maps for the region.  

In a few specific cases it was necessary to fit several Gaussians
to the profiles, as it was the case, for example, when  measuring
the expansion velocity of a newly found bubble in the centre of 30
Doradus in section \ref{find_bubble}. For that, we used the code {\sc pan} \citep{dimeo05}   
to fit multiple (and also single) Gaussians to the observed
profiles. This code, which runs in {\sc idl}, allows the user to
define interactively multiple Gaussians, estimating the flux, width
and centre of each component.

\section {Results}
\label{Results}

In this section we describe the main kinematic features found
in the H$\alpha$ Fabry-Perot data cube of 30 Doradus. 

\subsection{H$\alpha$ map and flux calibration}
\label{fluxcalibration}

At first we attempted to create the monochromatic map of the 30
Doradus region by fitting a single Gaussian to each pixel. However, we concluded
that the fit of single Gaussians to the often-encountered multiple
H$\alpha$ profiles brought no meaningful information to the
monochromatic map, given the complex profiles of the nebula.  
The 2D collapsed data cube was then used as the H$\alpha$
monochromatic map (shown in the top panel of Figure \ref{mono_vel_sigma}), 
assuming that the
contribution from the spectral continuum is negligible (and we do
measure a very low continuum of less than 1\% for the whole region).  
This approach is also justified by the fact that we use an HST H$\alpha$ image 
(which includes continuum emission) for flux calibration (see below). 
 
In order to flux calibrate the monochromatic map of 30 Doradus, we have
used an archival HST image of this region, which was observed with
the WFC3 and using the narrow-band filter F656N.  We used the task
{\sc phot} in IRAF$^4$ to measure the instrumental fluxes of 31
regions located across the field (which were located inside the
same field of view of the SAM-FP H$\alpha$ image), inside an aperture of 3
arsec radius. These instrumental values were converted into physical
fluxes by using \textit{PHOTOFLAM} (which is the so-called inverse
sensitivity),
F=1.78462$\times$ 10$^{-17}$ erg$^{-1}$ cm$^{-2}$ s$^{-1}$ {\AA}$^{-1}$
and the FWHM of the filter of 13.9 {\AA}.
Details can be found in the website of the Space Telescope Science Institute\footnote{http://www.stsci.edu/hst/wfc3/phot\_zp\_lbn and \\
http://www.stsci.edu/hst/wfc3/documents/ISRs/2003/
\\
WFC3-2003-02.pdf respectively.}.
Finally, each measurement was divided by the
area within which the flux was estimated.  For the SAM-FP 2D collapsed
image, we obtained the number of electrons per second for the same
regions defined for the HST image, inside the same apertures, using
the same procedure (here we use the gain and exposure times listed
in Table  \ref{table1} to transform counts in electrons/s).  In Figure
\ref{correlation} we show the result of this analysis, where a
dashed-dotted line represents a linear fit to the data, with 
a slope of 1.32$\times$10$^{-15}$ erg cm$^{-2}$
electrons$^{-1}$ (we note that this value is consistent with the
slope derived by forcing the zero point to zero, 1.38$\times$10$^{-15}$
erg cm$^{-2}$ electrons$^{-1}$).  Then the
SAM-FP 2D image, i.e., the monochromatic image, was calibrated by using this
coefficient (as shown by the scale on the right side of the top 
panel of Figure \ref{mono_vel_sigma}). 

Most of the features visible in the monochromatic map of 30 Doradus have already been
studied in the literature.  For instance, the bright Northeast and
a portion of the West filaments described as ionisation fronts by
\citet{rub98} and \citet{pel10}  are clearly
outlined. Many dark clouds are easily seen in contrast against the
bright nebula, for example the ``stapler-shaped" dark cloud \citep{walborn13}.
Another structure seen is the dusty main
cavity or {\it Christmas Tree}, which was detected by previous
studies (e.g. Chu \& Kennicutt 1994, De Marchi \& Panagia 2014). 

\subsection{A brief description of the kinematics of the observed region}
\label{A brief description of the kinematics of the observed region}

In the middle and bottom panels of Figure \ref{mono_vel_sigma} we show the velocity field and the velocity 
dispersion map for the central
region of 30 Doradus.  
Single Gaussian fits were used for obtaining these maps, regardless the number of line components. 
In the velocity field the fit traces the velocity peak at each pixel, which gives a general view
of the kinematics of the nebula. In the case of the velocity dispersion map, 
the fit to a single Gaussian is very useful, even when multiple kinematic
components are present, given that the width of the line 
allows the potential detection of regions with multiple kinematic components.
The velocity dispersion map was corrected by instrumental and
thermal widths, $\sigma_{inst}$ =11.3 km s$^{-1}$, $\sigma_{th}$ =
9.1 km s$^{-1}$, respectively. The latter value was estimated by
assuming an electronic temperature of T$_{e}$=10$^{4}$ K in the
expression $\sigma$$_{th}$=(k~T$_{e}$/m$_{H}$)$^{1/2}$.  

One of the most intriguing features of the velocity field of 30 Doradus (middle panel of Figure \ref{mono_vel_sigma}) is the central region of the main cavity, or {\it Christmas Tree} (R.A $\sim$
05h~38m~50s, Dec $\sim$ -69d~05m~52s).  Around this location there
is an abrupt change in the values of radial velocities, of $\simeq$
40 km s$^{-1}$, suggesting the existence of a complex kinematic
structure. This map shows that the kinematics in 30 Doradus
changes in small spatial scales.
The structure of the velocity field poorly correlates with the
H$\alpha$ distribution, in contrast with the velocity dispersion
map, which strongly correlates with it.  In the velocity field, we observe a global
velocity gradient of $\simeq$ 80 km s$^{-1}$ from Southeast to North and
Southeast to West. In addition, a component which either mimics or
is a real continuous velocity component is observed from Southeast
to Northwest (at $\simeq$ 295 $\pm$ 10 km s$^{-1}$). 

Inspecting the bottom panel of Figure \ref{mono_vel_sigma},
we note that 
regions having the largest values of $\sigma$ are
those for which two or more emission line profiles
are superimposed at different velocities.  This map is then a useful
tool to search for expanding structures.
We find that most of the high surface brightness regions (brighter than 
5 $\times$ 10$^{-15}$ erg s$^{-1}$ cm$^{-2}$) display narrow profiles ($\sigma$$\sim$20-30 km s$^{-1}$)
while low-intensity regions display broad multiple profiles
(70-80 km s$^{-1}$), which arise from unresolved emission line
components.  In some cases the emission line profiles are clearly resolved in several peaks, specially in the region of the 
{\it Christmas Tree}.  This area displays abrupt
changes in the number of components, intensity and radial velocities
of the ionised gas, which could be a result of the strong extinction along the line of sight.   

\begin{figure*}
 \includegraphics[width=\linewidth]{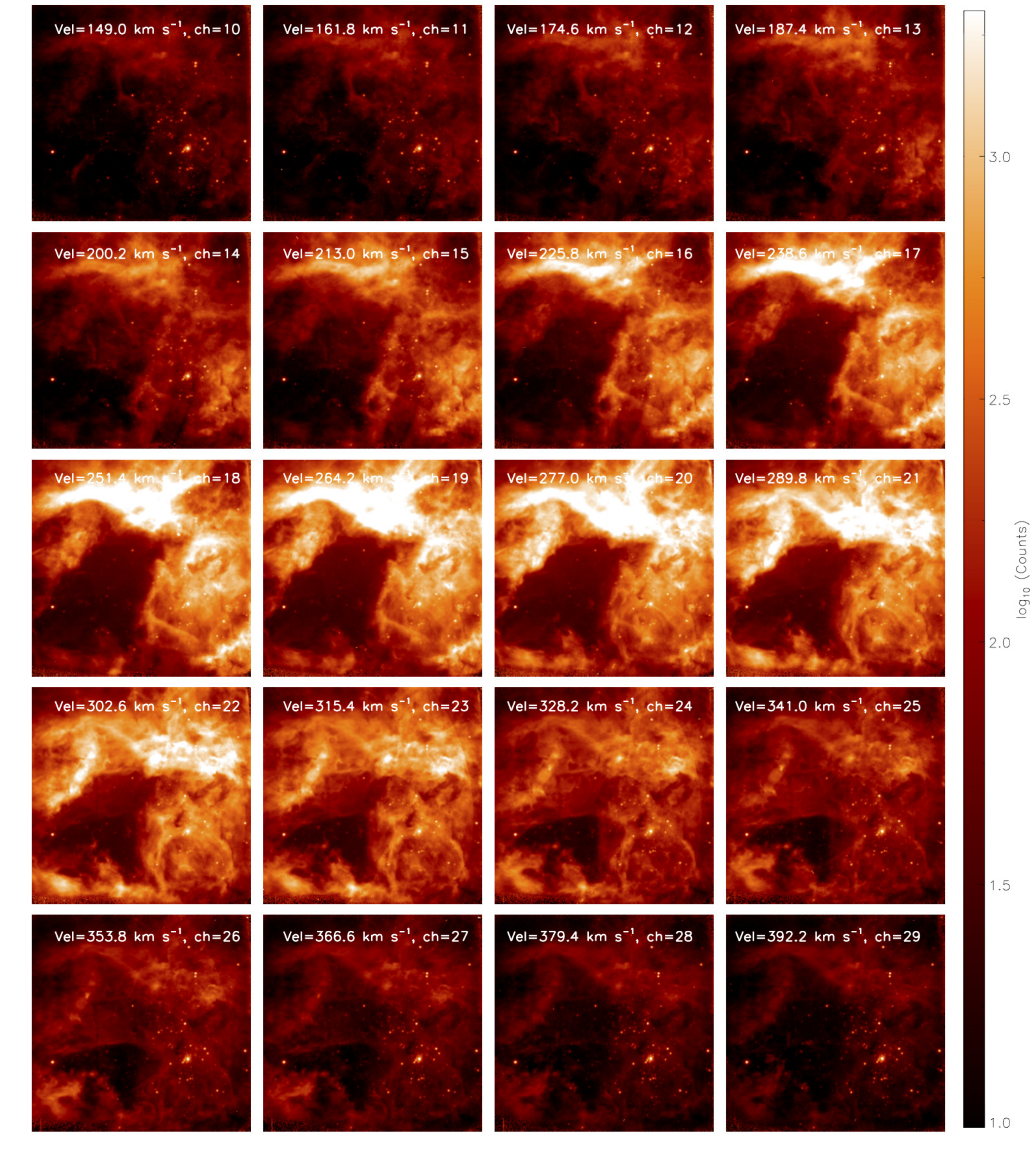}
 \caption{Channel maps for the H$\alpha$ data cube of 30 Doradus. We show here the 20 channels (of the 40 channels observed) with most emission. 
The Northeastern filament is the bright emission on the upper parts of the frames (seen in its brightest emission from channel 16 to channel 21). 
The {\it Christmas Tree} can be seen, in most channels, 
as the dark feature in the Southeast of the frames. The mean velocity of the channel and the channel number are shown at the top of each stamp.}
\label{hi_1}
\end{figure*}

In order to better display the kinematics of the inner region of
the 30 Doradus Nebula, in Figure \ref{hi_1} we show the more relevant
(in terms of flux) channel maps of the H$\alpha$ data cube (channels
10 to 29, 149.0 to 392.2 km s$^{-1}$).  
Starting from channel 16 (225.8 km s$^{-1}$) we can see some emission
associated with blue shifted gas. Between channels 18 (251.4 km s$^{-1}$)
and 20 (277.0 km s$^{-1}$) we
have the peak of the emission. The cavity or {\it Christmas Tree}
clearly appears in several channels.  This cavity progressively disappears
in channels 24 to 26 (328.2 km s$^{-1}$ to 353.8 km s$^{-1}$). 
 
\begin{figure}
\begin{centering}
\includegraphics[width=0.95\linewidth]{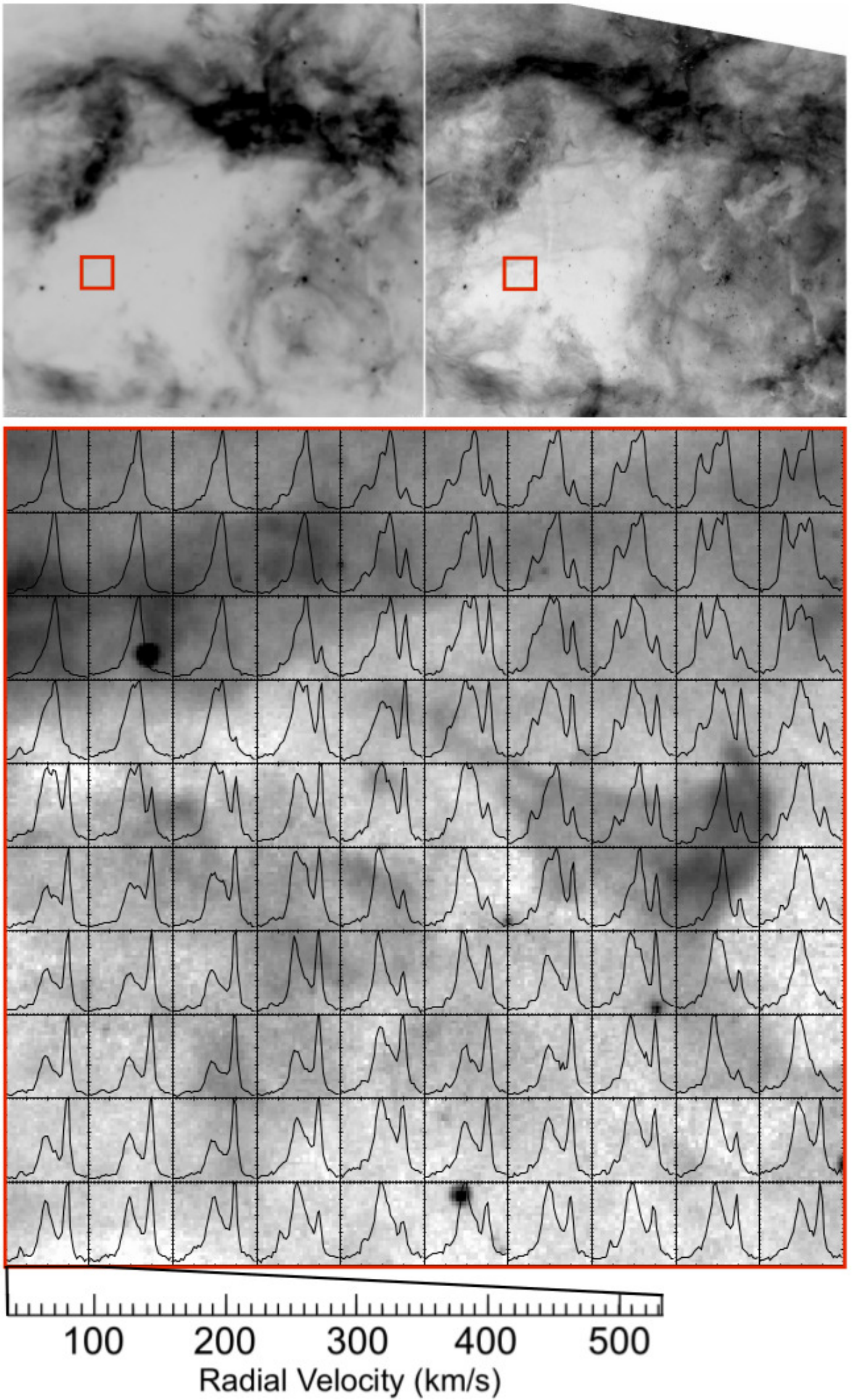}
 \caption{
 Profiles of one selected area of 30 Doradus. Top left pannel: H$\alpha$
image extracted from channel 21 of the Fabry-Perot data cube (see 
Figure \ref{hi_1} for a color version of this channel). The
red square indicates the region whose profiles are shown in the
lower panel. Top right panel: HST H$\alpha$ image of the same
region displayed on the top left. Bottom panel: H$\alpha$ profiles
over a 5 $\times$ 5 array of binned pixels. Each ``pixel" is 0.9 arcsec
(0.23 pc) on a side. The purpose of this Figure is to show the diversity of
profiles in a small area of only 9 $\times$ 9 arcsec$^{2}$ (0.23 $\times$ 0.23 pc$^2$). This diversity is
typical of the whole area observed. Note that 
the profiles can be easily traced from one pixel to the
next. The velocity scale is shown on the lower left, in one case, and it is the same for all profiles.
}
\label{profiles}
\end{centering}
\end{figure}

Perhaps the most striking result as we inspect the cube of 30 Doradus is
the diversity of profiles on the smallest scales.
This is exemplified in Figure \ref{profiles} for an area of
9 $\times$ 9 arcsec. Here, each square corresponds to a profile binned 
5 $\times$ 5 pixel$^2$, i.e, over a 0.9 $\times$ 0.9 arcsec$^2$ area
(which corresponds to 0.23 $\times$ 0.23 pc$^2$).  
The profiles change abruptly between neighbouring boxes, with components 
appearing and disappearing in scales of a fraction of a parsec (down to the spatial 
resolution of 0.15 pc). 

A detailed analysis of the kinematics of the whole  
2'$\times$2' field, using tools such as the 3D spatio-kinematic code SHAPE \citep{ste06} to 
disentangle the 3-D geometry and kinematic structure of 30 Doradus, is deferred to a future paper.

\subsection{Identification of a new bubble in the centre of 30 Doradus}
\label{find_bubble}

\begin{figure}
 \centering
\includegraphics[width=0.47\textwidth]{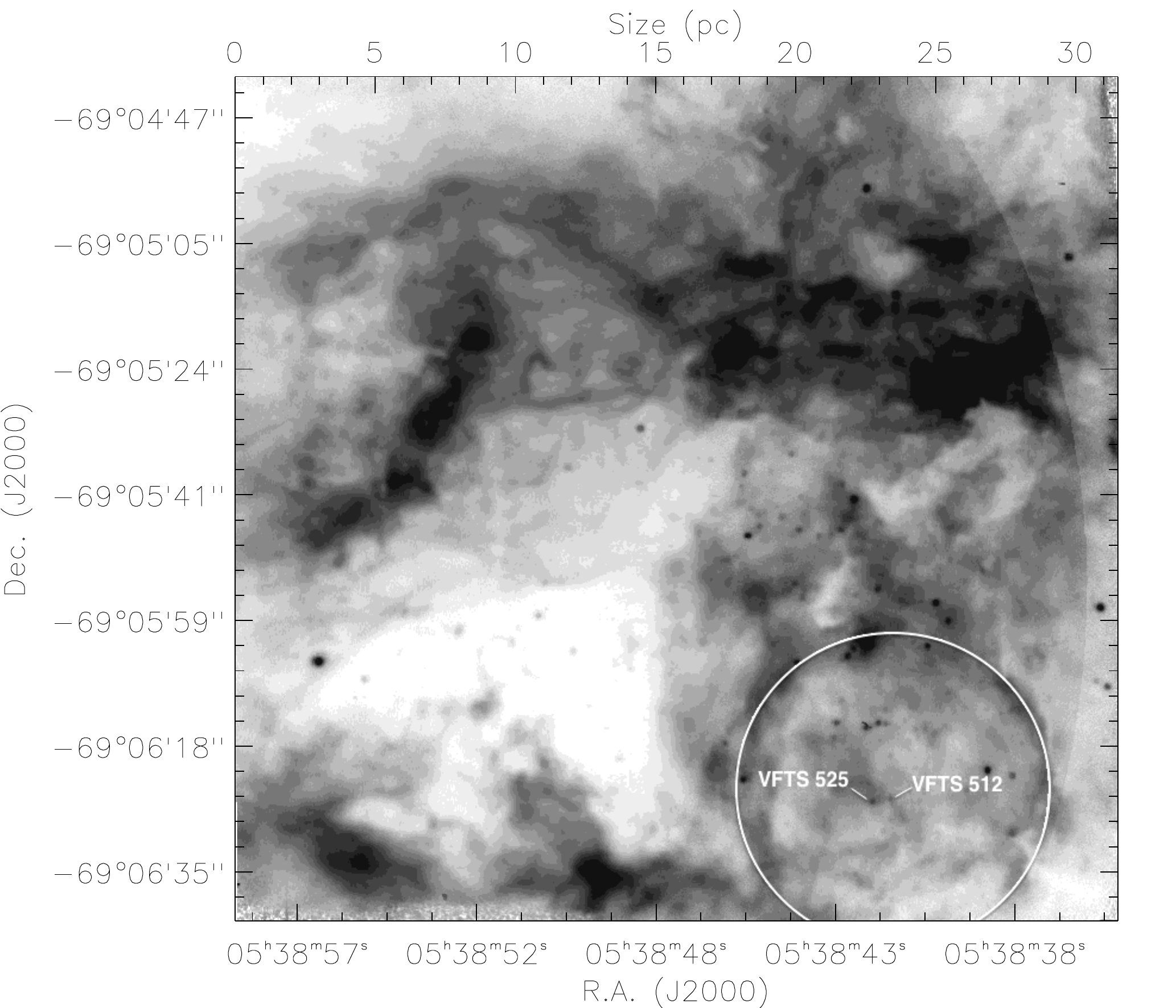}
\includegraphics[width=0.47\textwidth]{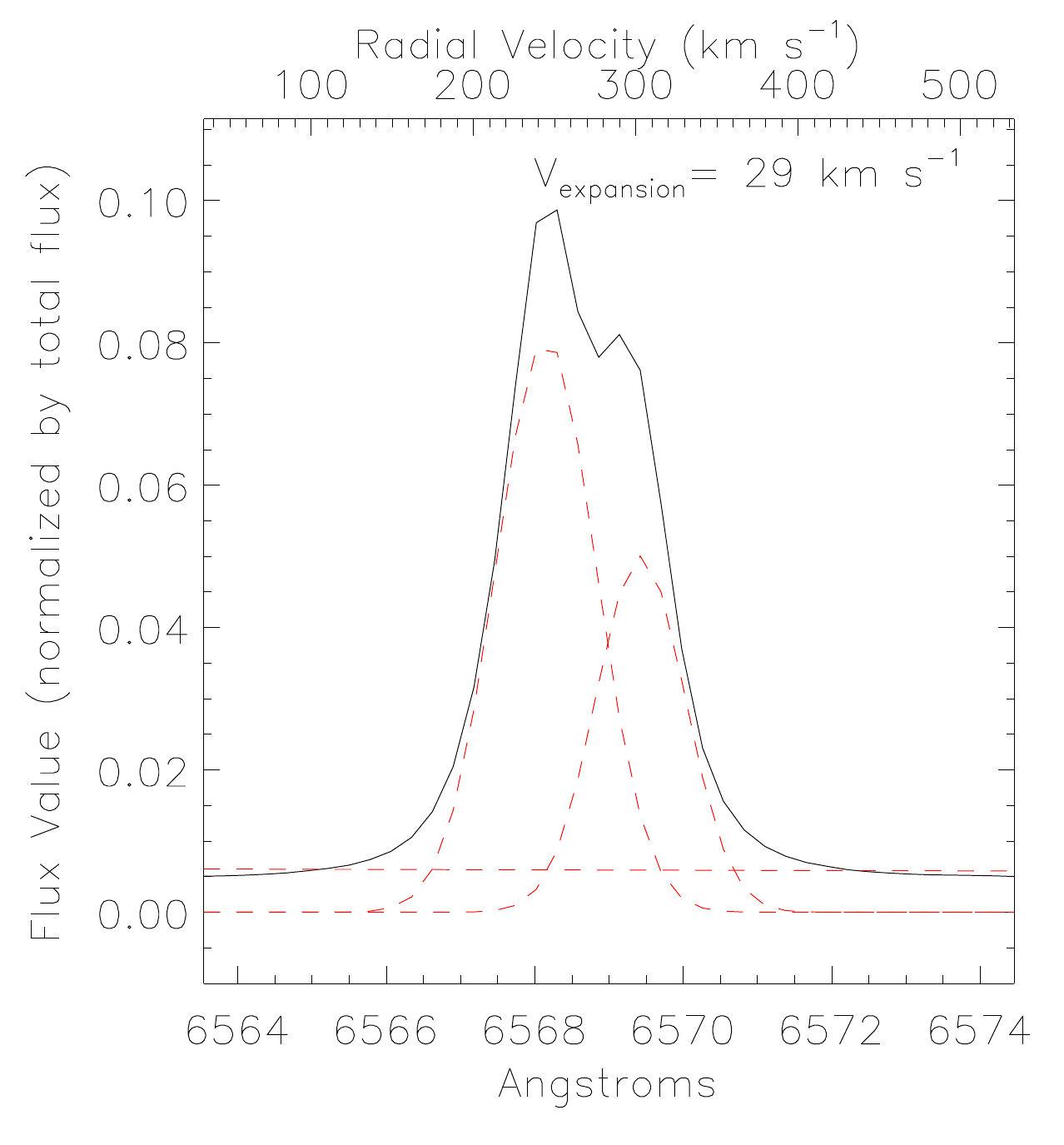}
\caption{Top panel: channel 23 (315.4 km s$^{-1}$) of the 30 Dor
data cube, where the white circle indicates the location of the
expanding bubble. The blue supergiant Melnick 38 and the very massive
spectroscopic binary VFTS 512 are indicated.  Lower panel: integrated
H$\alpha$ profile for the region contained within the white circle
shown in the top panel. Red dashed lines represent two Gaussian
components plus a continuum component, which can be identified as
a horizontal line, with a value of $\sim$ 0.006. On the top right we show
the value of the expansion velocity derived from the average of the 
velocities of the two peaks (receding and approaching sides).  
  } \label{expanding}
\end{figure}

The good spatial coverage and resolution of SAM-FP together with the
reasonably good spectral resolution allows identification of new
small-scale kinematic structures, which can not be easily identified
in previous sparse multi-fiber data or in long-slit studies, 
where the positioning
of the fibers and slits can bias the
detection of such structures. 
Given that the aim of this
paper is to show the capabilities of SAM-FP, here we show an example
of an expanding structure, which appears when inspecting the H$\alpha$
data cube. From the channel maps shown in Figure \ref{hi_1},
one can see the presence of an expanding bubble identified here
for the first time in the lower right of 
channels 20 to 25
(277.0 km s$^{-1}$ to 341.0 km s$^{-1}$). 
The presence of this bubble is not obvious in the velocity field
of 30 Doradus (middle panel of Figure \ref{mono_vel_sigma}),
but it can be identified as a region of broader profiles (i.e. 
larger widths - when using a single
Gaussian fit) in the velocity dispersion map (bottom panel of the 
same Figure).  We attempt to show this bubble more clearly 
in the top panel of Figure \ref{expanding}, where a slice of the
Fabry-Perot data cube centred at velocity 315.4 km
s$^{-1}$ (channel 23) is displayed.  The white circle indicates the location of
the expanding bubble centred at R.A $\sim$ 05h~38m~42s, Dec $\sim$
-69d~06m~23s. 

\citet{torresflores13} identified a large bubble that encompasses
this region (their region 8) and also includes the central cluster
R136. However, they had a poor spatial coverage of one spectrum
each 20 arcsec. With the present dataset we can more precisely
identify the location and size of this expanding bubble.  In Figure
\ref{expanding} (top panel) it is possible to see arc-like features, which
delineate the structure, approximately following the contours of
the white circle. 
The kinematic signature of this source is
shown in the bottom panel of Figure \ref{expanding}, which is the
result of adding all the H$\alpha$ profiles that are enclosed inside
the white circle (top panel). In this plot, we can see a double-peaked
profile, which is typically associated with expanding structures.
In order to quantify the expansion velocity of this structure, we
have fitted two Gaussians to the profile (plus a low-intensity
continuum emission). This procedure allows us to derive the radial
velocity of both components. Under the assumption that these
components are associated with the approaching and receding sides
of the expanding bubble, the expansion velocity of the bubble can
be estimated as V$_{\rm{expansion}}$= (V$_{\rm{receding}}$-V$_{\rm{approaching}}$)/2.
This exercise yields a velocity of V$_{\rm{expansion}}$=29  $\pm$ 4 km s$^{-1}$.
The radial velocity of the barycentre of the bubble,
(V$_{\rm{receding}}$+V$_{\rm{approaching}}$)/2, is 273 km s$^{-1}$. The error of 4 km s$^{-1}$
is obtained from the velocity difference between the barycentre measured on the observed double-peak 
profiles and the barycentre derived from the sum of the two fitted
Gaussians. Considering
that the bubble has a projected radius of 22 arcsecs or 5.6 pc,
we can speculate that this structure has been expanding
for a period of $\sim$ 180,000 years, assuming a constant
and uniform expansion.

The ionising source (or sources) associated with this
expanding bubble cannot be easily identified using FP data alone,
given that the continuum emission is poorly sampled in the data-cube.
Inspecting the population of massive stars in the central 10 $\times$ 10 arcsec$^2$ 
(or 2.5 $\times$ 2.5 pc$^2$ box) of the proposed bubble,  we have two remarkable stars, indicated in Figure \ref{expanding}:
the blue supergiant Melnick 38 (VFTS 525, B0 Ia, Walborn et al.
2014) and the very massive spectroscopic binary VFTS 512 (O2 V-III,
Walborn et al. 2014). These stars are located 2.7 arcsec and 2.4 arcsec from the
centre of the bubble, respectively, and they have different reasons to be the
source of the expanding nebula. In the case of VFTS 512, we could
be witnessing the birth of a pristine young bubble produced by a
powerful wind and UV-radiation of a very young massive star. 
In the case of Melnick 38, the origin of the nebula could
be related to advanced evolutionary stage of a blue supergiant.
\citet{walborn97}  described that in 30 Doradus at
least five distinct populations of massive stars are coeval, from an embedded
population of massive stars (see also Rubio et al. 1998, Walborn,
Barb\'a \& Sewilo 2013), the Orion Nebula-like population, to a more
evolved sparse population, composed by blue and red supergiants, the
Sco-Cen OB association-like population. \citet{naze01} searched
for pristine bubbles in two very young H{\sc ii} regions of the LMC, 
N11 and N180.
They discovered a very small bubble around the multiple star PGMW
3120, with a composed spectral type of O5.5V((f)), which is described
as a very young asymmetric bubble by \citet{Barba03}.

The new expanding bubble found in this work is
located close to the star cluster R136 (its centre 
is 5.6 arcsec South of R136, in projection).  
Although the mechanical feedback of R136
stars could disrupt any small-scale kinematic structure located in
its environment, this does not seem to be the case here, given the
apparently smooth shape of the bubble contours and its distinct
kinematics.  Probably this bubble is not located immediately next to R136
and the spatial coincidence we see is just a chance
alignment, in the line of sight.  Projection effects should be taken
into account in the identification and analysis of bubbles.

\subsection{The integrated H$\alpha$ profile}

\begin{figure}
 \includegraphics[width=\linewidth]{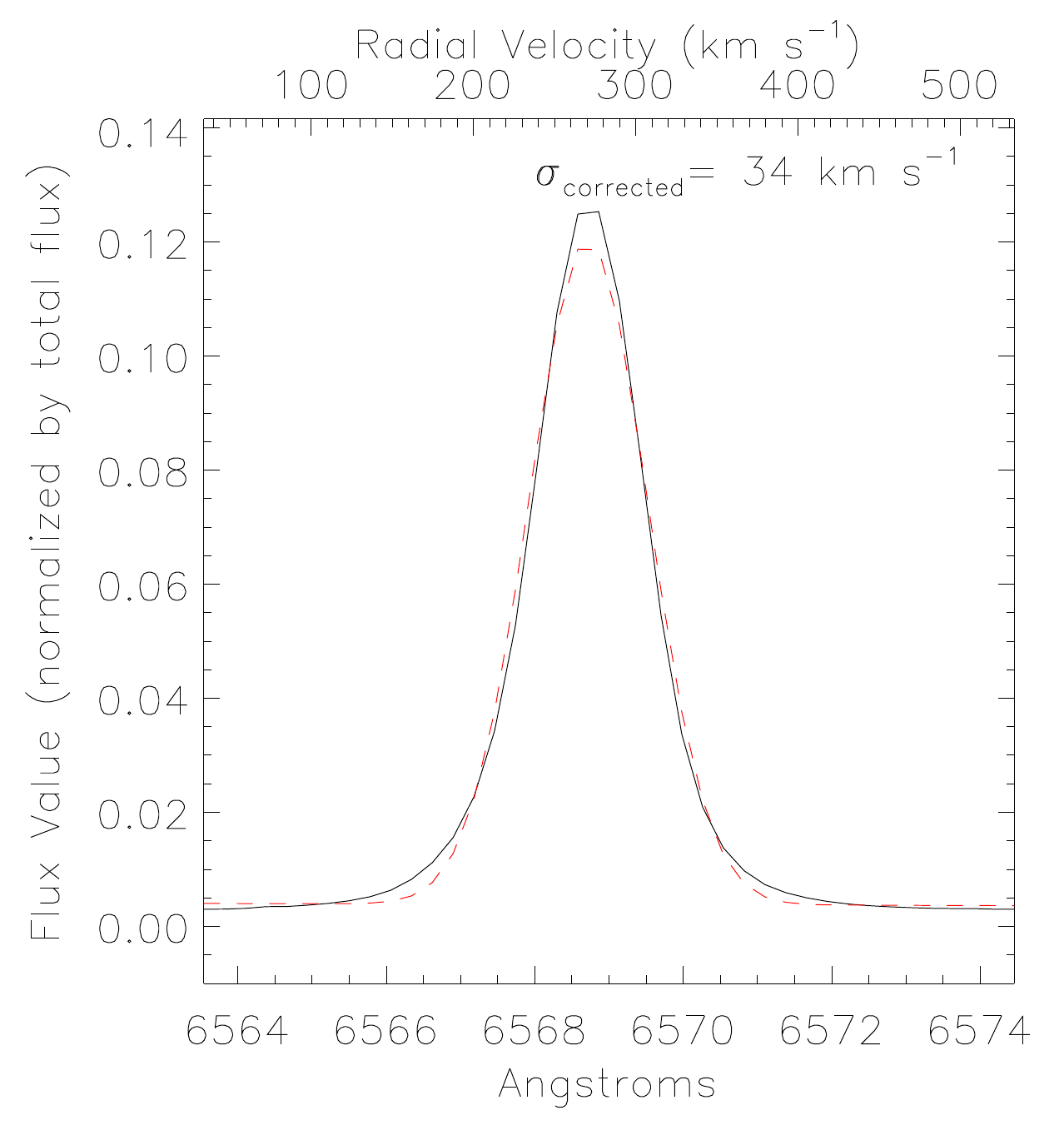}
\caption{Integrated H$\alpha$ profile of 30 Doradus, derived from
the whole data cube (black continuous line). The red dashed line
corresponds to a single Gaussian fit to the observed profile.
On the top left corner we list
the value of the width ($\sigma$) of the integrated profile, which
has been corrected by instrumental and thermal widths.} 
\label{integrated}
\end{figure}

A few studies to date have suggested
that the integrated H$\alpha$ profile of 30 Doradus is composed by a broad component, due to a low-intensity
emission,  superimposed on a narrow, high-intensity component. This may 
be caused by the sum of expanding structures or may be the result of the superposition
of a number of narrow, low-intensity components with
different radial velocities. We expect the broad component to be quite faint and therefore
a detailed and careful analysis and modelling of the line profiles, taking into account the underlying intrinsic Fabry-Perot Airy function,  needs to be done.
While this problem is outside the scope of the 
present paper, it is still interesting to measure the width of the integrated emission 
line of the whole region observed, to compare with that measured in the previous work of \citet{torresflores13}.
For this, we stacked all profiles of the 30 Doradus cube (a 2 x 2 arcmin$^2$ region). We then fit a single Gaussian to
the integrated H$\alpha$ profile, which is shown in
Figure \ref{integrated}. The black continuous line in this Figure corresponds
to the integrated observed profile, while the red dashed line is the
modeled single Gaussian (which was obtained by using the code {\sc pan}
in {\sc idl}).
We corrected the profile for instrumental and thermal 
broadening, in a similar way done for the velocity dispersion map and
we then computed its dispersion to be  $\sigma \sim $ 34 $\pm$ 1 km s$^{-1}$. 
The error of 1 km s$^{-1}$
is obtained from the velocity difference between the barycentres of the observed profile
and of the Gaussian fit.
The value obtained here for the dispersion is higher than the
one found by \citet{torresflores13} of $\sigma$=26.5 km
s$^{-1}$, but that was for a much larger region of 10'$\times$10'
centred on R136. These two values are consistent, given the uncertainties
due to different resolutions and different sizes of the regions
used in the determination.

In a forthcoming publication (Torres-Flores et al. in preparation) we will 
take into account the shape of the Fabry Perot profile (Airy Function)  
to correctly investigate the presence of wings in the integrated H$\alpha$ 
profiles, after bringing all different radial velocity profiles
to a common zero point velocity. 

\subsection{Visualizing the H$\alpha$ data cube in 3D}

3D visualization of the H$\alpha$ Fabry-Perot data cube of 30 Doradus
can help us understand the structure and kinematics of the nebula.

\begin{figure}
\begin{center}
 \includegraphics[width=0.40\textwidth]{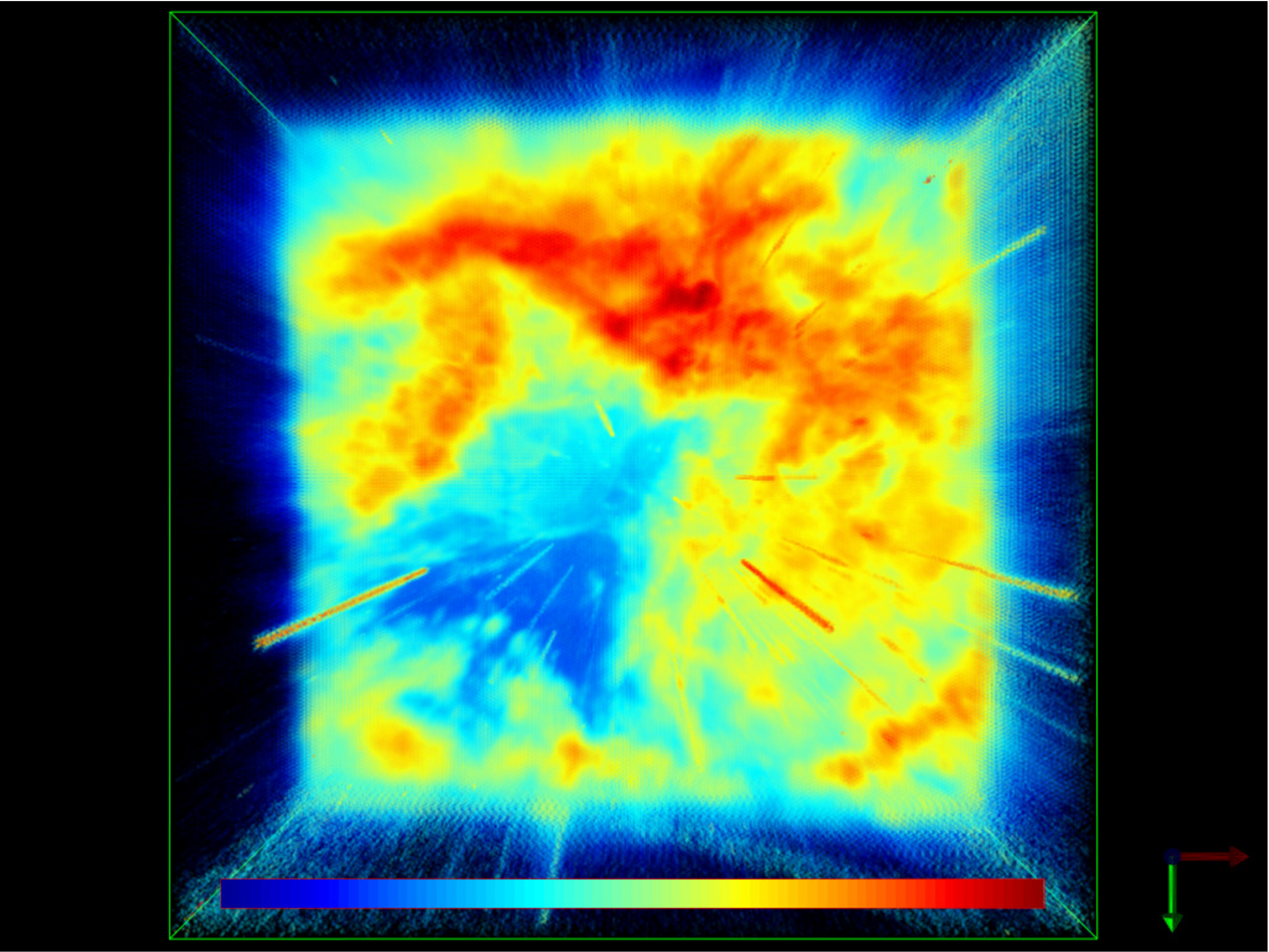}
 \includegraphics[width=0.40\textwidth]{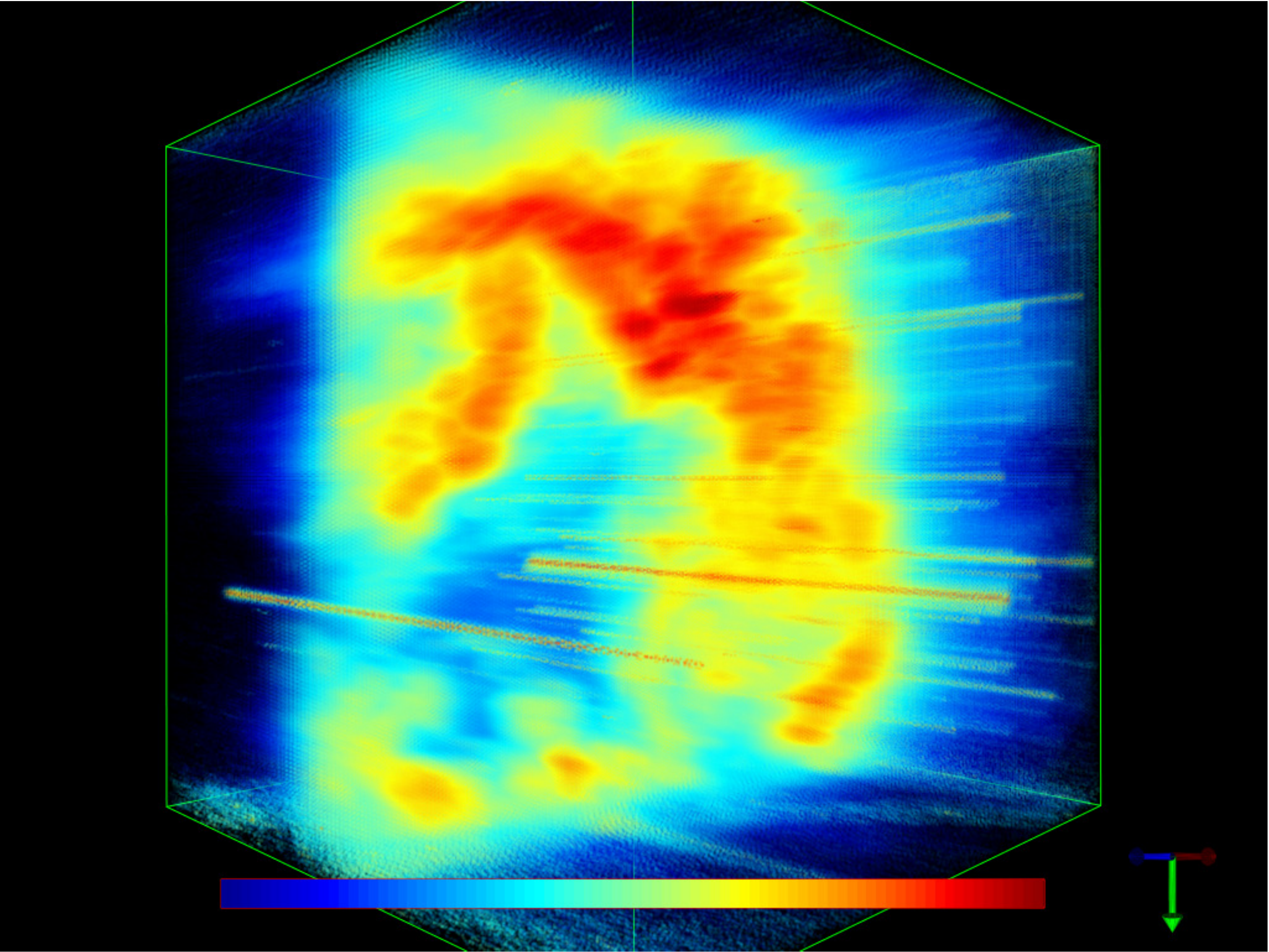}
 \includegraphics[width=0.40\textwidth]{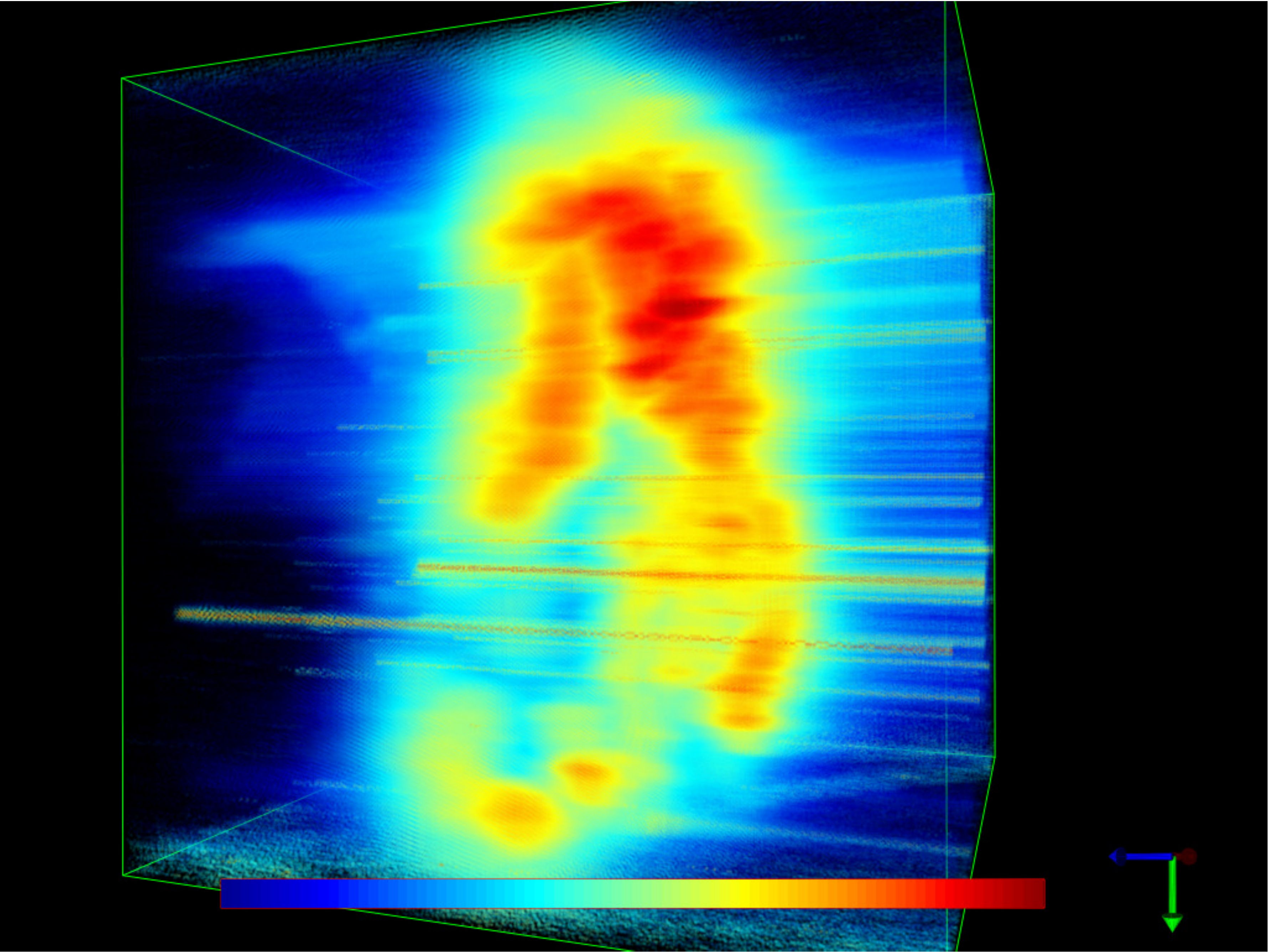}
 \includegraphics[width=0.40\textwidth]{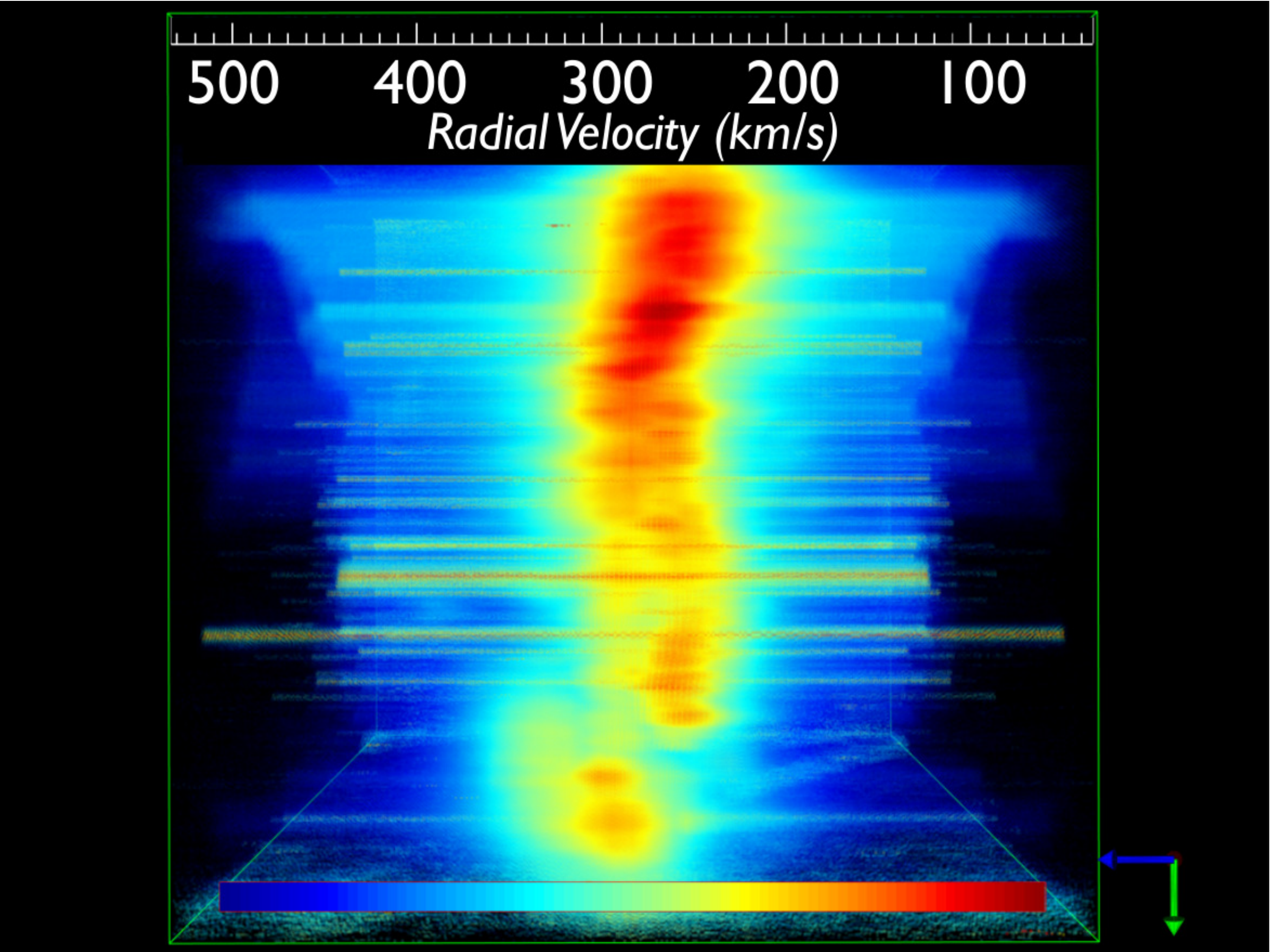}
  \caption{
3D view of the H$\alpha$ Fabry-Perot data cube of the
central part of 30 Doradus with rotation
around the declination axis.  Top: frontal view of 30 Doradus;
second from top to bottom: 45-degree view; third from top to bottom: 70-degree view; bottom:
90-degree view. The color bar shows a relative intensity scale. A velocity scale is given at the top of
the lowest panel. Continuous lines are stars in
the field. This Figure was prepared with the software GLnemo2 to facilitate
visualisation.}
\label{GLNemo}
\end{center}
\end{figure}

Figure \ref{GLNemo} shows one such 3D representation, where right
ascension, declination and velocity (wavelength) of 30 Doradus are
plotted.  In order to highlight small scale kinematic features, the
wavelength axis has been resampled to 350 channels (instead of the
original 40 channels). The panels of Figure \ref{GLNemo} were produced with 
GLnemo2 (Lambert 2016), an 
interactive 3D program that allows 
visualising the data cube in different
directions. These are indicated by arrows in the right-hand
corner of each panel, where the red, green and blue arrows correspond
to right ascension, declination and velocity (wavelength), respectively.
Depending on the angle along which the data cube is seen, we can obtain
from a face-on (0-degree view) image to a position velocity map
(90-degree view).
Figure \ref{GLNemo} shows a 3D view with rotation around the
declination  axis, with a frontal view, then a 45-degree, a 70-degree
and a 90-degree view of the system. Regions of greatest intensity
are shown in red while blue regions are those with lowest intensity and
the straight lines are the stars, which help constructing the 3D
view.

The contours of the cavity or {\it Christmas Tree}
are very well defined in the
first three panels of Figure \ref{GLNemo} (all except the 90-degree panel), given the contiguous
coverage of emission over the field.  In the top, a frontal view is shown,
where several low-luminosity structures can be seen within the cavity,  which 
are thought to be filaments being ejected in the line of sight.  They 
were seen by both \citet{melnick99} and \citet{chu94} in 2D images 
and long-slit spectra as numerous gas blobs
moving with more than 100 km s$^{-1}$ with respect to the system.  This
suggested to these authors that this is a region where expanding
wind-driven shells undergo break-out.  
These kinds of structures
were also identified by \citet{redman03} in another region of 
30 Doradus. In this sense, a detailed analysis of  H$\alpha$ SAM-FP
data will be extremely valuable in order to determine the main
mechanism producing these discrete knots.

The different radial velocities
across the nebula shown in the 90-degree cut (the position-velocity
diagram) also indicate the existence of several kinematic structures,
possibly parts of one or more expanding shells.

Another remarkable feature of Figure \ref{GLNemo} is the clear
indication that the Northeast filament (Rubio et al. 1988, the
filament that runs horizontally on the top part of the Figure) has
a different velocity than the Eastern filamentary loop (also identified as
ionisation front 2, or ``if2", Pellegrini et al. 2010). 
If we measure the
difference in velocity between the channel where the Northeast
filament is brightest and that where the Eastern loop is brightest,
we find 37$\pm$13 km s$^{-1}$  (we consider the error to be $\pm$ one channel, 
or $\sim$ 13 km s$^{-1}$).

There is a clear velocity gradient of the Northeastern filament,
with increasingly more negative velocities running from West to
East.  This can be clearly seen in the position velocity
diagram (bottom panel). Note that in this figure the Northern part of the
nebula, in projection, is clearly bent to the right, indicating
blue shift with respect to the bulk of the gas. As described by
\citet{Westmoquette10}, when a gas cloud hits a molecular cloud,
it flows around it and we see only the components that are facing
towards us (blueshift), given that the components that go behind
the cloud are screened by the cloud itself.  This process may be
happening in the Northeastern filament, with gas clouds coming
from the Southwest and hitting a large molecular cloud detected
by several previous works (e.g. Rubio et al. 1988 and references
therein). This process which here appears to happen in a scale of
25 pc or larger scales has also been suggested by \citet{Westmoquette10}
to occur around the Eastern pillar of NGC 6357 in a much smaller
scale.

The preliminary analysis described above shows clearly that the detailed kinematic 
map derived for complexes and filaments of 30 Doradus
may allow identification of several kinematically independent structures.  

\section {Summary and Conclusions}
\label{summary}

We present the first results obtained with a Fabry-Perot mounted
inside SAM, a ground-layer adaptive optics instrument using laser,
on SOAR.  The data cubes had a spectral resolution of $ R \simeq 11200 $ at
H$\alpha$ and a FWHM of $\simeq$ 0.6 arcsec.  This configuration provides a
unique tool to study the kinematics of a variety of astronomical
sources, such as normal and interacting galaxies, merging objects
and Galactic objects such as HH objects, planetary nebulae, and Giant
H{\sc ii} regions, with good spatial and spectral resolution. 
Note, however, that some targets, which have velocity ranges larger than the FSR of 492 km/s, are not appropriate for SAM-FP when used with the Fabry Perot device described in this paper. FPs with larger FSR must be used, for example,  to observe some planetary nebulae fragments, novae shells and supernova remnants, which can exhibit much faster velocities (e.g. the knots of MyCn 18, Redman et al. 2000, or nova GK Pe, Liimets et al. 2012) that will not be  captured by this velocity range. Moreover, objects with broad lines (> 8 \AA), such as active galactic nuclei, have to be observed with an FP with a larger FSR. In a future paper we will describe a second Fabry Perot that may be useful in such cases.

 An integral part of this paper is the material presented in the appendices, which provide recommendations on how to use Fabry-Perot instruments and how to understand the observations, providing all formulae involved. A few examples of topics discussed in the appendices are given in the following. A long standing confusion of observers about the difference between the scanning sampling and the resolution is clearly explained in appendix A, while the phase-shift effect, which is a physical effect linked to the coating of the etalons is described in appendix B.   We clarify when and why the observer may choose to scan a full FSR or a fraction of it or more than a FSR and we give some practical information for performing observations, such as for example, we recommend that the observer obtain a new observation in a position free of an object, for sky subtraction, when the object is larger than the field of view.  Without this knowledge, no proper observations can be done or the observed data will not be scientifically useful. Thus, the appendices are meant to be useful not only for users of SAM-FP but users of Fabry Perot instruments in general.
 
In order to highlight the capabilities of SAM-FP, we observed
the region of 30 Doradus, for which there are previously published
kinematic data.  We show an overall kinematic description of the
central part of the 30 Doradus nebula in the LMC. The current
configuration allows us to obtain kinematic information on  scales
of 0.15 pc, which can be improved under better seeing conditions.

Our main conclusions on the study of 30 Doradus are:

Comparing our Fabry-Perot data with previously taken VLT/FLAMES observations, 
we find good overall agreement in the regions of overlap, but the
SAM-FP data cubes have a filling factor 360 times higher than the 
comparison data cubes.

The Fabry-Perot dataset has great advantages over other multi-fiber
or long-slit data in the search for bubbles in nebulae in general
and in 30 Doradus in particular, given its much improved spatial coverage.
Analysis of the Fabry-Perot data reveal a new bubble at a projected
distance of 22 arcsec South of R136
with an expansion velocity of 29 $\pm$ 4 km s$^{-1}$.

The Eastern filamentary loop of 30 Doradus is redshifted by $\sim$ 37 $\pm$ 13
km s$^{-1}$ with respect to the Northeastern filament indicating
that these complexes are kinematically independent.

A velocity gradient in the Northeastern filament increases to more
negative velocities running from West to East.  This suggests the
presence of gas flowing around the molecular clouds, where we only
see the component facing towards us.  A similar mechanism was invoked
in smaller scales around the Eastern pillar of NGC 6357
\citep{Westmoquette10}

Small high-velocity clouds are seen inside the cavity known as
{\it Christmas Tree}.  These are also the same kinds of phenomena reported
by \citet{redman03}, where the shell is fragmented by perturbations
(such as small molecular clouds) in the medium.

In a future paper we will focus on the analysis of the whole data
cube and we will specifically study the ionised gas structure around
the massive stars, the relationship between their spectral types,
evolutionary stage and the kinematics of the H$\alpha$ profiles associated
to them. The aim is to identify a number of pristine H{\sc ii} regions
and small bubbles around massive stars and to evaluate the evolutionary
scenario of the bubbles. In this context, we will attempt to describe
the elusive kinematics of the small bubbles in the caldera of a
giant H{\sc ii} region (only possible with the data taken in this work).
Finally, this new instrument opens the possibility of exploring the
synergy between HST and FP data given that the wealth of details
seen in HST images of nearby giant H{\sc ii} regions can now be matched
by similar quality kinematic data taken from the ground, with SAM-FP
and laser correction.

\section{acknowledgments}

We warmly thank Marco Bonatti for writing the control software for
scanning the Fabry-Perot inside SAM. We thank Brian Chinn for
designing the mechanical mount for including the Fabry-Perot in
SAM. We thank Dr. Warrick Couch, Dr. Matthew Colless and the AAO
for kindly loaning us the Fabry-Perot instrument used for this work.
We thank Dr. Andrei Tokovinin for extensive help with the observing
runs and for very useful comments on the manuscript.  We are grateful
to  Dr. Tiago Ribeiro for taking the image of 30 Doradus used in
Figure \ref{field} with the T80-South telescope and to the T80-South
team for allowing us to use the image in this paper. We greatly
appreciated the help by Luciano Fraga, who agreed to exchange one
of his nights with ours, allowing tests and science observations
to be done with SAM-FP.  We thank Dr. Beno\^it Epinat for fruitful
discussions on Fabry-Perots, Mohamed Belhadi and Amandine Caillat
for measuring the passbands of the interference filters in several
configurations and Jean-Charles Lambert for facilitating the use
of GLnemo2.  We are greatly in debt to Dr. Keith Taylor, Rene Laporte
and INPE, for continuous support over the last many years to the
BTFI and SAM-FP teams. We thank Dr. Christof Iserlohe, for writing
the routine fluxer, which was used in this work.  We thank an
anonymous referee who made very useful suggestions to the original
version of the manuscript.  CMdO and DA thank FAPESP and CNPq for
financial support through grants 2009/54202-8 and 2014/07586-2.
ST-F acknowledges financial support from the ``Direccion de
Investigación y Desarrollo de la ULS'', through a project ``DIULS
Regular'', under contract PR16143. RB acknowledges financial support
from FONDECYT, through Project 1140076.  ST-F and CMdO acknowledge
financial support of CONICYT + PAI/Atracción de Capital Humano
Avanzado del Extranjero, Folio Nº PAI80160082.  CMdO and PA thank
the support of the USP-Cofecub international collaboration program,
which made this project possible. In particular the BTFI instrument
described in Mendes de Oliveira et al. (2013) and the papers
Torres-Flores et al. (2013b, 2014) and Alfaro-Cuello et al. (2015)
were possible thanks to the research exchange program supported by
USP-Cofecub, to which we greatly acknowledge.  Based on observations
obtained at the Southern Astrophysical Research (SOAR) telescope,
which is a joint project of the Ministerio da Ci\^encia, Tecnologia,
e Inova\c c\~ao (MCTI) da Rep\'ublica Federativa do Brasil, the
U.S.A. National Optical Astronomy Observatory (NOAO), the University
of North Carolina at Chapel Hill (UNC), and Michigan State University
(MSU).

\appendix

\section{What is needed for carrying out Fabry-Perot observations}
\label{What is needed for carrying out Fabry-Perot observations}

This appendix is written as a recipe. First we describe how to select the calibration lamp and the interference filter. Then we give the list of parameters that need to be provided/computed to obtain Fabry-Perot observations. Finally some recommendations on how to perform the observations are given. Section 2.2 of the main text and Appendix \ref{What is needed for understanding Fabry-Perot observations} explain why these steps should be followed.
  
\subsection{How to choose the right calibration lamp?}
\label{How to choose the right calibration lamp?}

The calibration lamp must contain at least one emission line passing through an available interference filter.  In order to minimize additional phase-shift effects, the interference filter for the calibration should ideally be the target interference filter (defined here as the interference filter suitable for the observation of the on-sky target).  However, often no narrow emission lines are available within the target interference filter, so the observer should select a calibration-lamp narrow emission line as close as possible to the observed emission line (H$\alpha$ in our case, redshifted due to the velocity of the 
target).   Around H$\alpha$, the 6598.99 \AA\ Neon emission line is suitable for nearby objects and a filter, centred, e.g. on 6600 \AA, with 20 \AA\ width, can then be used.  Further explanations are given in Section \ref{Wavelength calibrations and additional phase-shift effect}.

\subsection{How to choose the right interference filter?} 

Optical Fabry-Perot instruments are typically optimised to work in the blue or in the red, depending on the plates' coating. The observer should further narrow down the spectral range of interest by selecting the target interference filter. The FWHM of the target interference filter should match and in practice should be slightly larger than one FSR. The FSR (formula \ref{lfsr}) depends on the wavelength and on the interference order and it is typically 11 \AA\ for the Fabry Perot described in this paper. Thus the target interference filter could have a FWHM between 12 and 25 \AA (note that if the filter width is too large, then different orders of a given wavelength will appear in different
parts of the data cube causing possible confusion due to line overlapping). 
The right interference filters have to be chosen for both the calibration and object observations. 
Further explanations are given in section \ref{Interference filters}.

\subsection {Which input parameters are needed for performing FP observations?}

\subsubsection{How to compute the optimal number of steps?}

The computation of the optimal number of steps $\rm{n}$ needed to scan a FSR is described in this section.  To satisfy the Nyquist-Shannon's sampling criteria, $\Delta e_A$ must be sampled at least on two channels. 
In order to minimize the number of scanning channels, the Airy function is then sampled with
two channels only and thus the number of channels needed to scan a FSR is simply $n = (2 \times F$), recalling that F is the effective finesse of the etalon.
Given that F is a float number and $\rm{n}$ is an integer, then $n = (2 \times F$) + 1.
For instance the ICOS ET-65 etalon used for this work has an effective
finesse $F \simeq 18.5$, thus $n=38$. 
The observer is nevertheless
free to choose another number of steps $\rm{n}$ in order to undersample
or oversample the scanning sequence but of course it will not change
the resolution $R_{\lambda} $. This is indeed what we did, in the
case of the observations of 30 Doradus, 
we gently oversampled the scanning of the effective finesse by
10\% leading to $n=40$   
scanning channels and to a
scanning step of 12.8 km s$^{-1}$.  The determination of the FSR and of the effective finesse are given in sections \ref{Measuring the FSR} and \ref{Measuring the Finesse} respectively.  

\subsubsection{How to choose the scanning wavelength?}

The scanning wavelength for the object must correspond to its mean radial velocity. The scanning wavelength for the calibration must be the wavelength of the
calibration line.

\subsubsection{Which range to scan?}

The observer may chose to scan a fraction of the FSR, exactly the FSR or more than one FSR.  The first option is not typically recommended for high-order
Fabry-Perot instruments because the whole field of view will not be covered by the FSR.  The third option could be selected if the observer needs to check the FSR during the reduction procedure.  To optimize the observing time we recommend the second option, i.e., to scan exactly one FSR. Further technical details on the way to best define the scanning sequence is provided in 
section \ref{Defining the scanning sequence}.

\subsubsection{How to choose the observing time?}
\label{How to choose the observing time ?}

The description below assumes that the Fabry Perot instrument uses a camera with a high-quality CCD detector, which is the case in most large observatories. 

First of all, the observer has to determine the observing time per individual channel.  This time is a compromise between a reasonable number for the total time spent on target at a given stable atmospheric condition and the optimization of the use of the detector (given its read noise characteristics).  Cosmic ray hits also impose an upper limit to the exposure time per channel (to a maximum of about 5 minutes).  As an example, for a scanning sequence with 40 steps, the exposure per channel should not be longer than 2 minutes, given that this leads to more than 80 minutes for the total cycle, which is already a significant time during which the atmospheric conditions may have changed (airmass, transparency and seeing).  A shorter exposure time per channel is not advisable for low-surface brightness objects, given the typical readout noise of modern CCDs ($\sim$ 3 electrons). In case the observer wants to increase the signal-to-noise ratio of the observation, several cycles should be repeated.  Much shorter exposure times per channel (say 1-30 seconds) are suitable for high surface brightness sources (e.g. bright planetary nebulae and bright HII regions).

\subsubsection{How to choose the pixel binning?}

It is advisable to use binned pixels to match Nyquist-Shannon's
sampling criteria, avoiding high spatial oversampling.  Binning
will considerably increase the SNR per unit surface and this is
mandatory for low-surface brightness objects, given the readout
noise of the CCD and the typically short exposures used in Fabry-Perot observations.

\subsection{Standard observing procedures}

\subsubsection{About wavelength calibrations}

Once the scanning sequence is defined, the observer can use it for the wavelength calibration.  We recommend that lamp calibration frames are taken during day time to check the Fabry-Perot set-up and the scanning procedure (in this case the scanning wavelength must correspond to 
zero velocity, i.e. rest wavelength).  As explained in section \ref{takingcalibrations}, it is recommended that the observers do wavelength calibration before or after the target exposure with the telescope in the same position to match as much as possible the conditions of the observations.

\subsubsection{About subtraction of the night sky-line emission}
\label{How to get rid of the night sky line emission ?}

If the target does not fill the whole field of view of the detector, the night sky line could be safely derived and subtracted during the reduction procedure (using the 
galaxy-free area to measure the sky lines).  If the target does fill the whole field of view, two cases may be envisaged.  First, if the spectral density is low (i.e. the average wavelength range of the target emission line per pixel is less than about one third of the FSR), the night sky line could be identified using medians during the reduction procedure.  Second, if the spectral density is  high, it is recommended to measure the sky from a new pointed observation 
on a target-free location, just beside the target.  As the sky is expected to not have spatial structure, a short exposure should be enough (typically one tenth of the exposure time used for the target). After wavelength calibration, spectra from all pixels can then be summed up. 

\subsubsection{About flux calibrations and flat fields}

Flux calibrations for Fabry-Perot observations should be done using standard procedures like any other instrument, i.e. by observing calibration standards. We nevertheless recommend to use spatially extended calibrated emission line sources like planetary nebulae or isolated HII regions. A scanning sequence with the same parameters used for the target observations must be used (with the important difference that the scanning wavelength must be the wavelength of the calibration source, i.e., in the case of a planetary nebulae it should correspond to the radial velocity of the planetary nebula).  Flat fields should be taken to correct the spectrum from pixel to pixel efficiency variations and the effect of the interference filter transmission curves.  The Fabry-Perot should not be in the beam during the procedure of obtaining flat fields (so no scanning is necessary).  

\section{What is needed for understanding Fabry-Perot observations}
\label{What is needed for understanding Fabry-Perot observations}

To complement the main text and Appendix \ref{What is needed for carrying out Fabry-Perot observations}, this appendix gives the relevant explanations to understand Fabry-Perot observations. 

\subsection{Wavelength calibrations and additional phase-shift effect}
\label{Wavelength calibrations and additional phase-shift effect}
 
Before obtaining on-sky observations one must take a few reference calibration cubes for checking the Fabry Perot parameters and for subsequent wavelength calibration. This is usually done in day time, prior to the first night of observations. As described in Appendix \ref{How to choose the right calibration lamp?}, one must select an arc line (from one of the calibration lamps available at the telescope) which has a wavelength which is similar to that of the emission line to be observed.  

Selecting a proper arc line to be used for wavelength calibration is a crucial step for any Fabry Perot observation due to the additional phase-shift problem explained below.  In textbooks on Fabry-Perots, it is stated that the transmission curve of a Fabry-Perot is not a function of the wavelength within a given spectral band. However, in practice, the reflectivity of the coatings deposed onto the inner side of the two plates do depend on the wavelength. Each reflection, due to each dielectric coating, having complex transmission and reflection coefficient numbers, induces an additional phase-shift $\psi(\lambda)$, also called phase-lag, which 
depends on $\lambda$ in the range $[0,2\pi]$. Then equation (\ref{necosipl}) becomes:
\begin{equation}
 \rm{p = \frac{2ne \cos \theta}{\lambda}-\frac{2\psi(\lambda)}{2\pi}},
\label{ppsy}
\end{equation}
where p is now the effective interference order, $ \rm{n (T=0^oC, P=1\  atm)\simeq\ 1.0003}$ is the index of the
air layer between the two plates of the Fabry-Perot device, $\rm{e}$ the
inner separation between the coated plates and $\theta$ the incidence angle. 
What happens is that the coatings, which consist of the superposition of multiple 
thin dielectric layers (typically 0.1-1 $\mu m$ each), alternating transparent dielectric 
material with low refractive index (typically n$\simeq$ 1.5) and high
refractive index (typically n$\simeq$ 2.5), are meant to provide
an uniform electric field penetration depth within a given spectral
range but this aimed uniformity is never perfectly achieved. This
then induces the additional phase-shift that should be accounted for. The
way to minimize the effect of the additional phase-shift is to have the wavelength of the arc line 
be as close as possible to the wavelength used for the scanning of the science data cube.  If we
observe H$\alpha$ close to its rest wavelength (which is the case
for galactic H{\sc ii} regions, inter-arm regions, planetary nebulae,
SNR, or galaxies belonging to the Local Group for which systemic
heliocentric radial velocities are low or even negative), the ideal
situation would be to use a Hydrogen lamp to select the H$\alpha$
emission as the calibration line, but in practice this is not a good
option. The H$\alpha$ emission
line is intrinsically quite broad and it is then better to choose a narrower
emission line close to the wavelength of H$\alpha$ redshifted to the velocity
of the target.  As explained previously, 
the Ne 6598.95 \AA\ line (if a Neon lamp is available) can be
successfully used as a reference
line, given its intrinsic 
narrow shape and high intensity, allowing
accurate measurements.  For redshifted objects like nearby galaxies,
this Ne line is very well suited because it is close to the observed
redshifted H$\alpha$ line ($\lambda$=6598.95 \AA\ corresponds to the
wavelength of H$\alpha$ for an object with a
radial velocity of $\sim 1650 $ km s$^{-1}$). We note
that, in order to take observations with a low-interference order
Fabry-Perot, usually called a tunable filter, several lines in 
one calibration spectrum 
are needed to cover the broad FSR but for a high-order Fabry-Perot instrument,
which has a small FSR, one calibration spectrum with one spectral line may be enough (if a second calibration spectrum
is available one can then check the order). 

\subsection {Interference filters}
\label{Interference filters}

\subsubsection{Fabry-Perot transmission function}
\label{Fabry-Perot transmission function}
In order to complement the basic Fabry-Perot formulae
(\ref{necosipl}, 
\ref{lfsr},
\ref{efsr} and
\ref{spectralresolution})
given in section \ref{computingparameters}, we must introduce the Airy function. A Fabry-Perot provides a periodic signal described by the Airy function:
\begin{equation}
\mathscr{A(\phi)}= \rm{\frac{1}{1+(2F/\pi)^2\sin^{2}(\phi/2)}}
\label{airy0}
\end{equation}
where $\phi$ is the total phase-shift (or phase-lag):
\begin{equation}
 \rm{\phi=\frac{2\pi}{\lambda}\delta-2{\psi(\lambda)}}
\label{airy2}
\end{equation}
$\phi$ results from the combination of the optical travel path $\delta$: 
\begin{equation}
 \rm{\delta = 2 n e \cos i}
\label{airy3}
\end{equation}
and $\psi(\lambda)$, the additional phase-lag  described in section \ref{Wavelength calibrations and additional phase-shift effect}; F is the effective finesse; $\rm{n}$ is the refractive index of the medium between the two plates;  $\rm{i}$ is the beam inclination and $\rm{e}$ is the plate separation. 
The transmitted intensity $I_{t}$ of the Fabry-Perot is thus:
\begin{equation}
 \rm{I_{t}=I_t^{max*}\frac{T_o^2}{(A_o+T_o)^2}\mathscr{A(\phi)}}
\label{airy1}
\end{equation}
with
\begin{equation}
 \rm{A_o+R_o+T_o=1}
\label{ART}
\end{equation}
where $ \rm{I_t^{max*}}$ is the maximum transmitted intensity; A, R and T, the absorption, reflectivity and transmissivity coefficients respectively.  Note that $ \rm{I_t=I_t^{max*}}$ when $ \rm{\phi=2p\pi}$, which corresponds to the ideal case where absorption $\rm{A_o}$ (and diffusion) could be neglected in the interfaces; in this case, formula \ref{airy2} can be written as formula \ref{ppsy}. Note also that the additional phase-lag can be as large as the main phase-lag $2\pi \delta /\lambda$ when e and p are small, i.e. for a low-order Fabry-Perot, also called tunable filter.

\subsubsection{Suitable interference filters}

Considering formulae \ref{airy0}, \ref{airy2},  \ref{airy3} and \ref{airy1}, and as mentioned in Section \ref{computingparameters}, in order to select one or few orders, an interference filter must be chosen both for the target observation and for the calibration. In the case of the target observation, the central wavelength of the interference filter should correspond approximately to the mean velocity of the source. Depending on the science case, the observer can choose a filter allowing to select more than one FSR and make $a posteriori$ correction but, in practice, more than three FSR may result in great confusion of orders, especially for galactic sources. In addition, all the continuum emission selected by the interference filter passes through the Fabry-Perot and it is modulated by the transmission curve of the filter. Thus, in order to increase the contrast of the Fabry-Perot (i.e. the monochromatic detection power), the interference filter should be as narrow as possible so that the continuum flux passing through the Fabry-Perot is minimised, which is an important point given that the photon noise $\rm{\sqrt{N}}$ of a continuum emission will limit the monochromatic detection.  Moreover, in order to lower the continuum modulation, the passband of the filter should be as square as possible.

\subsubsection {More on interference filters}
\label{More on interference filters}

The observer has to take into account that the filter passband is blueshifted when 
(i) the incidence of the incoming light is not normal to the filter, (ii) the temperature drops and (iii) the filter ages.  In addition, often under these circumstances the filter peak transmission is reduced and the FWHM is enlarged.

For small angles, the peak wavelength of the interference filter is a function of the beam aperture, temperature and time and it can be described by:
\begin{equation}
 \rm{\lambda_{c} \simeq \lambda_{beam} -\alpha(\lambda_o) (T_o-T) - A(t)}
\label{filtre}
\end{equation}
where,
\begin{equation}
 \rm{\lambda_{beam} \simeq \lambda_o \sqrt{1-\left(\frac{n_o}{n_{f}}\right)^2\sin^2 (\theta+i)} }
\label{filtre2}
\end{equation}
Here, $\lambda_{c}$ is the corrected central wavelength of the interference filter and $\lambda_o$ the central wavelength at normal incidence; $n_o\simeq 1$ and $n_{f}$ are the air refractive index and filter effective refractive index respectively; $\theta$ is the beam aperture (e.g. the focal ratio at SAM focus is $f/16.63$ providing $\theta \simeq 1.7^o$) and $\rm{i}$ is the filter inclination; $T$ and $T_o$ are the ambient and reference temperature respectively, $\alpha(\lambda_o)$ a coefficient that slightly depends on the wavelength and finally $A(t)$ a coefficient that depends on the ageing of the filter and must 
be measured over a long enough period.\\

The three cases (i), (ii) and (iii) mentioned above will be discussed in the following. 

(i) Narrowband interference filters are thin and solid Fabry-Perot interferometers usually operating in first order. The air gap is replaced by a thin layer of dielectric material (also called cavity) designed with an optical thickness optimized for the desirable transmission wavelength. Multiple-cavity filters provide steeper band slopes (thus square passbands) by alternating layers of high-index material (typically Ta$_2$0$_5$, $ \rm{n_H} \simeq$ 2.13) and low-index material (typically SiO$_2$, $ \rm{n_L}\simeq$1.46). The effective refractive index of an interference filter is typically $ \rm{n_f\simeq \sqrt{n_H\ n_L} \simeq 1.76}$.

We have measured in the lab at the Laboratoire d'Astrophysique de Marseille the central wavelength blueshift when inclining to a small angle $ \rm{0^o < i < 6^o}$ a set of $\sim$20 interference filters.  In order to experimentally measure  $ \rm{n_f}$, but specifically to determine an empirical and operational relation, we fit equation (\ref{filtre2}) to the data. We find $ \rm{n_f=1.72_{-0.22}^{+0.35}}$.  The values vary quite a lot from one filter to another despite the fact that the process of fabrication is the same for all filters. For convenience, in table \ref{inclination}, we tabulate values computed for the blueshift ($\lambda_{c} - \lambda_{o}$) using formula \ref{filtre2}, in two cases, for $\lambda_{o} $= H$\alpha$ 6562.78 \AA\ and $\lambda_{o}$ = H$\beta$ 4861 \AA, using $ \rm{n_f=1.72}$.

\begin{table}
\caption{Wavelength blueshift of interference filter transmission curves}
\begin{center}
\begin{tabular}{ccc}
\hline
inclination ($^o$) & \multicolumn{2} {c}{blue shift (\AA)}  \\
			   & H$\beta$ 4861 (\AA) & H$\alpha$ 6563 (\AA)\\
\hline
0 & 0 &0 \\
1 & -0.2 &-0.3 \\
2 & -1.0 &-1.3 \\
3 & -2.1 &-3.0 \\
4 & -3.8 &-5.2 \\
5 & -6.0 &-8.1 \\
6 & -8.5 &-11.6 \\
\hline
\end{tabular}
\end{center}
\label{inclination}
\end{table}%

In practice, the filter could be inclined to voluntarily shift the transmission curve to the blue, but the inclination should be small (typically lower than a couple of degrees) because it enlarges the passband and decreases the transmission. Moreover the shifts in wavelength do not increase linearly, becoming quite large beyond angles of 6 degrees. 

In formula \ref{filtre2}, the angle $\theta$ represents the aperture beam of the focal plane in which the filters are usually located.  Usually Fabry-Perots are located in focal reducers to lower the focal ratio and increase the field of view.   The aperture $\theta$ is furthermore a telescope or focal reducer-dependent parameter. In the case of SAMI, given that it is not a focal reducer, the focal ratio in the filter plane is the one of the telescope (f/16.63).\\

(ii) The ambient and reference temperature are usually different.  Manufacturers typically provide a reference temperature $ \rm{T_o=20^o}$C but the temperature when acquiring the observations is usually lower. To measure the temperature dependence, we used data provided by the company Melles Griot, Optics \& Photonics Company. From their data we obtained:
$$ \rm{\alpha(H_\alpha\ 6562.78 \AA)=\frac{\Delta \lambda}{\Delta T} = 0.24\pm\ 0.01 \AA\ K^{-1}}$$
Other manufacturers provide lower values of $\alpha$, ranging from 0.15 to 0.19 $ \AA\ K^{-1}$ in the R-band. Using the Melles Griot data, within the optical range, we computed a weak dependence of $\alpha$ with $\lambda$, 
$$ \rm{\alpha=0.175\ \left(\frac{\lambda}{\lambda_{H\alpha}}\right) + 0.063},$$ 
providing for instance $ \rm{\alpha(H_\beta\ 4861)= 0.19 \pm \ 0.01\ \AA\ K^{-1}}$.\\

(iii) The ageing effect on the wavelength (blueshift of the central wavelength) is very difficult to generalize because it affects differently each filter, even for a set of filters having the same age and built by the same company. It depends on the use and storage conditions. The filters measured in this experiment have, however, the same ``history" and we measure A(t$\simeq$15 years) $\simeq$ 3$\pm$ 3 \AA.  
Some filters even displayed a red-shifted passband but measurements done 15 years apart may have been realised with different setups. 

Moreover, transmission curves of old filters may completely collapse and their FWHM may dramatically increase. Peak transmissions and FWHMs of all filters that were 15 year-old were measured. In the worst cases, we find peak transmissions of 24-40\% and as broad as 31-45 \AA\ (for filters showing, when new, a peak transmission of 70\% and FWHM=20-24 \AA\ respectively).  For the whole sample, we measured a mean peak transmission of $54\pm16\%$ and a mean FWHM of 24 $\pm$ 9.7\ \AA\ to be compared with values, when new, of mean peak transmission of $71\pm2\%$ and mean FWHM of 17.6 $\pm$ 4.2\ \AA\ (the FWHM scatter when the filters
were new only reflects the fact that their initial widths ranged from 12 to 24 \AA,  but the increase on the scatter is significant). 
Here again, even if their process of fabrication, conditions of use and storage are similar, the scatter in peak transmissions and widths is very high. This means that some filters still exhibit good transmission curve even if they are 15 years old, while others that are just as old have low transmissions and they have been degraded with time. Thus, some low-transmission filters are clearly, in practice, not useful any longer. The conclusion is that filters must be regularly measured.\\

\subsection{Determining FP parameters}
\subsubsection{Determining the interference order p}
\label{Determining the interference order p}

Formula (\ref{necosipl}) gives the interference order p. Although a number for the interference order is given by the manufacturer of the Fabry-Perot, it  is also possible to measure it more accurately taking into account that the distance between the plates may vary slightly. This is, however, 
quite tricky, given that two different
methods are needed to measure the integer part of p, $ \rm{p_o=int(p)}$,
and its fractional residual, $ \rm{\epsilon=p-int(p_o)}$.  However, for our
purposes, the value of $ \rm{p_o}$ is the most important, which
simplifies the computation.  $ \rm{p_o}$ can be computed using a filter
and a lamp whose spectrum has two emission lines in the wavelength range of interest. 
$ \rm{p_o}$ can be obtained using these two lines (two different wavelengths) using the classical
method of coincidences.
One can find the interference order for $ \rm{p_0}$ by determining where in the field the interference rings coming from the two different wavelengths spatially coincide.   
Once this integer is known  $\epsilon$ is computed by measuring the radius
distribution of the parabolic rings.

\subsubsection{Measuring the FSR}
\label{Measuring the FSR}

In this section we explain how to measure the FSR for a
Fabry-Perot in wavelength and in the so called binary control value
(BCV, explained below), which is linked to the Fabry-Perot controller.
These procedures can then be generalised for almost all scanning Fabry Perots in use.

Formulas \ref{lfsr} and \ref{efsr} give the FSR in wavelength and as a function of the Fabry Perot plate separation, 
respectively.  In this section we will detail how to compute and measure it. 
It is important to measure the FSR before a given observing run (from calibration cubes) in order to minimize the observing time
spent on-object without losing any information.  The goal
is to do a scanning cycle on-object that corresponds exactly
to one FSR or it is just slightly larger.  If the scanning sequence
covers less than one FSR, some areas of the field-of-view will not
be observed and some wavelengths will be missing as well.  If the
scanning sequence covers x $\times$ FSR (with x $>$ 1), some (x $-$ 1)
interferograms are redundant and are useless because the new
incomplete cycle does not cover a whole FSR.  The FSR 
has to be 
measured at the calibration wavelength $\lambda_c$.  If $\lambda_c$
is close to the observed wavelength $\lambda_o$, the additional phase-shift
is negligible and, using equations (\ref{necosipl}) and (\ref{efsr}),
it is straightforward to compute the FSR at $\lambda_o$ from the
FSR computed at $\lambda_c$.  In practice, the determination of
the FSR is done by scanning
a gap slightly larger than the expected $\Delta e_{FSR} $ in order to get
more than two maxima of the cyclic Airy function.

If $ \rm{e_m}$ is the mean distance between the plates
around which the plates could be adjusted, and $\Delta e_{M}$ the
maximum separation physically reachable between the plates (limited by the
course of the piezoelectric actuators), the distance between the
plates of the interferometer varies in the range 
$ \rm{\sim [(e_{m} -\Delta e_{M})/2, (e_{m} + \Delta e_{M})/2]}$. 
If one wants to control 
the parallelism between the two Fabry-Perot plates and move one plate
with respect to the other, three degrees of freedom are needed: $X$ and $Y$ control
the parallelism between the Fabry-Perot plates and $Z$ controls the spacing within 
a certain dynamic range.
In our specific case, when using the CS 100 controller with the ICOS ET-65,
the $Z$ offset between the plates has a full dynamic range of 12-bit 
binary number (note that for other Fabry Perot controllers, this dynamic
range could be different - but this specific case is given here as an example). 
Thus, the maximum number of Binary Control
Values (BCV) is $ \rm{\Delta Z_{M}=2^{12}=4096}$, spanning within the
range [$ \rm{-Z_m=-2048, Z_{m}-1 =+2047}$]. $ \rm{\Delta Z_{M}}$ corresponds to
approximately 2 $ \mu $m of plate separation adjustment. For the
CS100 controller that drives the ICOS ET-65 etalon, the range of
$ \rm{\Delta e_{M}/2 \simeq \pm1 \mu m}$ around $ \rm{e_m}$ can be reached using
a $\pm 10V$ differential input, with a non-linearity of the scan
of $\pm$ 1\% and with a response time of 1 ms at a frequency response
of 160 Hz (3dB)\footnote{from ICOS, CS100 Controller and ET Series
II User's Guide.}. The smallest increment of plate separation
adjustment $\delta e$ is thus 

\begin{equation} 
 \rm{\delta e = \frac{\Delta e_{M}}{\Delta Z_{M}} }
\label{deltaeZ} 
\end{equation} 

Furthermore, the number of BCV needed to scan the full FSR, $ \rm{\Delta
Z_{FSR}}$, is linked to the corresponding increment of plate separation
to scan the FSR,  $ \rm{\Delta e_{FSR}}$, and to the smallest increment
of plate separation, $ \rm{\delta e}$, by:
\begin{equation} 
 \rm{\delta e = \frac{\Delta e_{FSR}}{\Delta Z_{FSR}} }
\label{deltaeFRS} 
\end{equation} 

$ \rm{\delta e}$ is the physical expansion of the piezo-stacks for 1 BCV
step applied to the etalon.  

The exact value of the
plate separation $\rm{e}$ and $ \rm{\delta e}$ are not known with high
accuracy. We need to determine them using e.g. the calibration
line $ \rm{\lambda_c}$.  In order to switch from the expression of the
FSR in terms of plate separation to the equation of the FSR
as a function of the wavelength, we use equation (\ref{efsr}) which gives $ \rm{\Delta e_{FSR} = \lambda/2 }$ at the centre of the rings ($\theta=0$) for $\rm{n}$ = 1. Substituting equation \ref{deltaeFRS} in the previous expression, 
we get $ \rm{2 \delta e = {\lambda}/{\Delta Z_{FSR}}}$, which provides the definition
for the ``Fabry-Perot scanning constant" Q:
\begin{equation} 
 \rm{Q = 2\ \delta e =  \frac{\lambda}{\Delta Z_{FSR}}}
\label{qcc} 
\end{equation} 

The ``Fabry-Perot scanning constant" Q is equal to twice the
smallest increment of plate separation $ \rm{\delta e}$, conventionally
given in units of \AA/BCV.  In the present case, for the ICOS ET-65 Fabry-Perot, 
equation (\ref{deltaeZ}) provides a rough value $ \rm{Q_{rough}}$ $\simeq$
2 $\times$ 1 $\rm{\mu m}$ / 2048 $\simeq$ 9.8 \AA/BCV.  The physical expansion
of the piezo-stacks for 1 BCV step $ \rm{\delta e}$ (thus Q) depends
on the electronics of the controller that commands and
controls the piezoelectric.
In addition, it varies slightly with temperature, pressure and
humidity and should be measured with accuracy. Several methods could
be used, the simplest is to start from a given interferogram (ring
image) and increase the BCV with very small steps, i.e. 
high sampling, to cover more (e.g. 10-20\% more) than the
expected FSR. A highly oversampled calibration cube, with spectral
coverage of more than one FSR, should be obtained in day time, ahead
of the first night of observation, for this purpose and for computing
the effective finesse. The
image for which the initial pattern is observed a second time
indicates that a FSR has been scanned and $ \rm{\Delta Z= \Delta Z_{FSR}}$.
For instance, applying this procedure to 
the  high resolution Fabry-Perot with $ \rm{ p \simeq 609}$ at Ne
6598.95 \AA, we measure $ \rm{\Delta Z_{FSR} \simeq 705}$ BCV (which
is $\sim$17\% of the maximum range $ \rm{\Delta Z_{M}}$).  This now allows
an accurate determination of Q using the relation \ref{qcc}: 
$ \rm{Q_{acc}= 9.360\pm0.003 \AA/BCV}$ at 6598.95 \AA.

\subsubsection{Measuring the effective finesse}
\label{Measuring the Finesse}

Formulae \ref{airy0} and \ref{airy1} provide the Airy function which depends on the finesse F. 
This section explains the nature of the finesse and how to measure it.  In textbooks on Fabry-Perots, the finesse F only depends on the reflectivity $ \rm{R_o}$ of the coating and could be computed
using the optical finesse  $ \rm{F_o}$ which is the mirror reflectivity finesse:
\begin{equation}
 \rm{F_o = \frac{\pi \sqrt {R_o}}{1-{R_o}}}.  
\end{equation}
In reality, Fabry-Perot interferometers differ in several ways from the ideal case and this impacts the actual finesse, that is enlarged.  This actual finesse is referred to as the effective finesse.  As described in section \ref{Wavelength calibrations and additional phase-shift effect}, the reflectivity $ \rm{R_o}$ and the transmissivity$ \rm{T_o}$ coefficients depend on the wavelength (similar to the phase-shift, upon reflection at the plate surfaces). Moreover, the coating and the plate substrates induce absorption (described by equation \ref{ART}), which also depends on the wavelength and reduces the transmitted intensity. Indeed, when the absorption $\rm{A_o}$ increases, following equation \ref{airy1}, the transmitted intensity decreases.  In addition, real Fabry-Perot interferometer plates present a variety of defects, they show irregularities at their surface, they are not perfectly parallel or planar and the plate separation $\rm{e}$ may vary due to instability in the control command of the Fabry-Perot.  Finally the effective finesse of the interferometer will depend on different factors if the Fabry-Perot is located in a collimated (parallel) or in a converging beam. The Fabry-Perot could be located in a pupil plane, in a focal plane or somewhere in between along the optical plane. If the Fabry-Perot is at (or close to) a pupil plane it is usually in a collimated beam and if the Fabry-Perot is anywhere else, and {\it{a fortiori}} at (or close to) a focal plane, it is in a converging beam.  In the most general case, the finesse F is the combination of the optical finesse $ \rm{F_o}$ and of the so-called control-command finesse $ \rm{F_c}$, imaging finesse $ \rm{F_i}$, aperture finesse $ \rm{F_a}$, and defect finesse $ \rm{F_d}$.  Assuming the finesse functions are close to Gaussians (which is an acceptable first order approximation), the effective finesse is then:
\begin{equation}
\rm{\frac{1}{F^2}=\frac{1}{F_o^2}+\frac{1}{F_c^2}+\frac{1}{F_i^2}+\frac{1}{F_a^2}+\frac{1}{F_d^2}}
\end{equation}
Depending on the location of the Fabry-Perot device along the optical path, this equation takes different forms.  Independent of its location, the optical finesse $\rm{F_o}$ and the control-command finesse $\rm{F_c}$  always play a contribution. $\rm{F_c}$ is due to control-command inaccuracies due to electronic instabilities in the plate separation positioning, mainly due to the piezoelectric actuators but also due to the gain and the time-constant adjustments. In other words, the controller provides a plate separation $\rm{e}$ with an associated dynamic instability error $\rm{\delta e}$ that affects the finesse. If the Fabry-Perot is placed in a converging beam, the beam aperture induces a shift on the transmitted wavelength as well as a decrease of the finesse. These two effects depend on the position in the field. Usually, the wavelength shift is the most important effect.
The aperture finesse $ \rm{F_a}$ decreases when the beam aperture $f/d$ decreases (where $f$ is the focal length and $d$ the diameter of the mirror).   In addition, $ \rm{F_a}$ also decreases when the interference order p increases, i.e. when the resolution increases, but it does not depend on the wavelength for a given p. It should be noted that the peak transmission (given by the equation \ref{airy1}) decreases as the beam aperture decreases, the effect being more pronounced the higher the order. On the other hand, the imaging finesse $ \rm{F_i}$ and the defects $ \rm{F_d}$ do not play a role because random point to point variations in the phase-shift between the plates do not affect the finesse in a converging beam (while it does in a collimated beam).

If the Fabry-Perot is placed at a collimated beam, then $f/d$ and $ \rm{F_a}$ tend to infinity, $ \rm{1/F_a^2}$ tends to zero and the beam intercepts the whole surface of the Fabry-Perot (limited by the pupil). However, the light rays are never strictly parallel, and this affects the image quality. However, even if the rays were parallel, the image quality is also dependent on the angular size of the seeing at the Fabry-Perot entrance. This is accounted for by the imaging finesse $ \rm{F_i}$ defined as:
\begin{equation}
 \rm{F_i=\frac{\lambda}{2ne \Delta \theta \sin \theta}},
\label{fi}
\end{equation}
where $\theta$ is the beam incidence on the Fabry-Perot and $\Delta \theta$, the image quality, i.e. the angular size corresponding to the seeing disk on the Fabry-Perot entrance.  $ \rm{F_i}$ depends on the illumination of the Fabry-Perot and, therefore, on the input beam diameter.  Note that for a given image quality, the imaging finesse depends on the plate separation and on the field-of-view. Finally, the defect finesse $ \rm{F_d}$ represents an important contribution to the effective finesse when the FP is at a collimated beam. Three potential defects may be associated with Fabry-Perot plates and are potential contributors to the instrumental broadening function \citep{Atherton81}: (1) the surface irregularities (roughness), coating deposit inhomogeneity on the plates; (2) the departures from parallelism (``wedging") between the plates and (3) the bowing of the plates due to mechanical stress because of the coating or the opto-mechanic interfaces. The defects mentioned in (2) and (3) also depend on gravity (i.e. on the position of the interferometer during telescope tracking).  These three defects account respectively for the surface quality finesse $ \rm{F_{d,s}}$, the wedging finesse $ \rm{F_{d,w}}$ and the bowing finesse $ \rm{F_{d,b}}$. Finally the defect finesse could be written as:

\begin{equation}
 \rm{\frac{1}{F_d^2}
= \frac{1}{F_{d,s}^2}+\frac{1}{F_{d,w}^2}+\frac{1}{F_{d,b}^2} 
= \frac{32 \ln 2 (\delta d_{s})^2+ 3 (\delta d_{w})^2+2 (\delta d_{b})^2}{\lambda^2}
}
\end{equation}

where $ \rm{\delta d_{s}}$ is the gaussian standard deviation (r.m.s) of the polishing plus coating errors, $ \rm{\delta d_{w}}$ the maximum wedge amplitude  (i.e. the peak-to-valley deviation from parallelism) and $ \rm{\delta d_{b}}$ the maximum (the peak-to-valley) excursion from the plane surface (i.e. the sagitta of the arc segment). All may cause blurring (broadening) of the line-shapes.   The slightest misalignment of the etalons'  plates can lead to measurements of the effective finesse and resolution that are slightly different from one run to the next, given that the system setup is fine-tuned for each run.

In practice, the effective finesse, F can be easily measured during the setup. This is
achieved by obtaining e.g. a scanning sequence at high sampling
also needed for computation of the FSR.  We use the formula:
\begin{equation}
 \rm{F =  \frac{\Delta e_{FSR}}{\Delta e_A}}
\label{finesse}
\end{equation}
where $ \rm{\Delta e_A}$ is the FWHM of the Airy function or, more precisely,
it is the increment in plate separation necessary to cover
one FWHM of the Airy function while $ \rm{e_{FSR}}$ is the
increment of plate separation necessary to scan the FSR.

Using equations (\ref{efsr}) and (\ref{finesse}) we find:
\begin{equation}
 \rm{\Delta e_A =  \frac{e}{pF}}.
\end{equation}
The average effective finesse for the whole field-of-view can be obtained by fitting an Airy function to the arc lines of each calibration spectrum, in the data cube. In practice it is sufficient to extract a 1D spectrum from a calibration data cube, e.g. in the centre of its rings (where $\theta=0$) and to fit an Airy function or simply measure the FWHM of an arc line.
F is then obtained by dividing FSR by the width of
the arc line. For the high-resolution Fabry Perot used for this study we
measure an effective finesse $ \rm{F \simeq 18.48}$ at Ne 6598.95 \AA\,
which gives $ \rm{ R \simeq pF \simeq  11260 }$ at Ne 6598.95 \AA\ and 
$ \rm{R \simeq 11200}$ at H$\alpha$.

\subsubsection {Defining the scanning sequence}
\label{Defining the scanning sequence}

This section describes how to define the scanning sequence. Once the
number of channels $\rm{n}$ is fixed and the Fabry-Perot constant Q
is known, the scanning sequence could be defined for a given
wavelength. Reference wavelength calibration data cubes are taken
at the calibration wavelength $\lambda_c$ (e.g. using the Ne I emission line
6598.95 \AA\ selected through a narrow interference filter $\sim$19
\AA) by scanning the FSR at $\lambda_c$.  A BCV value close to
Z = 0 (the middle of the plate separation adjustment) defines the
zero point of the scanning sequence that corresponds to half of the
FSR to scan.  
Nevertheless, other values different from Z = 0 can also be chosen as long as the FSR fits within the range
of BCV values (i.e., within $ \rm{\pm Z_m}$, e.g. at SAM-FP we regularly use Z = 2048 with the
scanning sequence going from 0 to 4095).
Using the equation  \ref{qcc}, e.g. at the calibration
wavelength $\lambda_c$, the first channel to scan thus corresponds
to the BCV:
$\rm{Z_1=-\Delta Z_{FSR}/2=-\lambda_c/2Q;}$
the scanning step between two successive channels is 
$\rm{\Delta Z_{FSR}/n=\lambda_c/nQ,}$
thus channel 2 corresponds to 
$\rm{Z_2=-\lambda_c/2Q+\lambda_c/nQ,}$
channel 3 to
$\rm{Z_3=\lambda_c/2Q+ 2\lambda_c/nQ}$ 
and the last
channel is 
$\rm{Z_n=-\lambda_c/2Q+(n-1)\lambda_c/nQ}.$
Ideally, we should plan to not have an overlapping of two successive
orders, thus channel $1$ and channel $n+1$ are identical. The first
and the $ \rm{(n+1)^{st}}$ channels match, as do also the second and the
$ \rm{(n+2)^{nd}}$ and so on. Finally, the step in BCV values between two
successive channels $ \rm{\Delta Z_{FSR}}$/n is an integer
value, thus 
$\rm{n \times \rm{nint}(\Delta Z_{FSR}/n)}$, where nint is the function nearest integer, is not exactly equal to $\Delta Z_{FSR}$.  The difference 
$ \rm{\left\vert \Delta Z_{FSR} -n \times nint \left( \Delta Z_{FSR}/n \right) \right\vert}$ 
must then be homogeneously spread over all steps of the scanning sequence.
This means that, in the end,
the step between any two successive channels could be one
BCV larger or smaller than the average step value. However, this is
not a problem given that the FSR is, in any case, scanned
with an accuracy better than one BCV (i.e. better than 0.14\% for
the setup used in this paper).

On-sky observations are carried out using the same scanning sequence
(number of channels) as for the calibration but at the wavelength
of the observation $\lambda_o$, in order to cover the whole FSR.
The number of BCV scanned is then not exactly the same as for the
calibration. Indeed, to cover a full FSR at the wavelength of the
on-sky observation $\lambda_o$, a new sequence should be defined by:
$$ \rm{\left[
-\frac{\lambda_o}{2Q},
-\frac{\lambda_o}{2Q}+\frac{\lambda_o}{nQ}, 
-\frac{\lambda_o}{2Q}+\frac{2\lambda_o}{nQ},...,
-\frac{\lambda_o}{2Q}+\frac{n-1}{nQ}\lambda_o\right]}
$$
The zero point of the scanning sequence ($Z=0$) should be common to all wavelength, i.e. the plate separation $\rm{e}$ of the Fabry-Perot should be the same regardless the wavelength because the equation 
$ \rm{p_c\lambda_c=p_o\lambda_o}$, deduced from the equation (\ref{necosipl})
for each point in the field of view (for a fixed $\cos \theta$), is only valid for
this plate separation $\rm{e}$. Consequently, if $ \rm{\lambda_o \lessgtr  \lambda_c}$,
$ \rm{\Delta Z_{FSR}(\lambda_o) \lessgtr \Delta Z_{FSR}(\lambda_c)}$.
Therefore, besides the additional phase-shift mentioned in the beginning
of this appendix, another reason to choose $\lambda_c$ as close as
possible to $\lambda_o$ is to minimize the difference between the
two FSR, minimizing the need for interpolation during the  construction
of the wavelength calibrated data cube. Nevertheless, here again,
the observer is free to use other scanning sequences, if desired.

\end{document}